\documentclass{aa}
\usepackage[varg]{txfonts}
\usepackage{rotating}
\usepackage{xcolor}
\usepackage[breaklinks=true, colorlinks, citecolor=blue, urlcolor = blue]{hyperref}
\usepackage{soul}
\usepackage{afterpage}
\usepackage[varg]{txfonts}
\usepackage{supertabular}
\usepackage[normalem]{ulem}
\usepackage{csquotes}
\usepackage{units}
\usepackage{mathtools}
\usepackage{float}
\usepackage{natbib,graphicx}
\usepackage{txfonts}
\usepackage[caption = false]{subfig}
\usepackage{amsmath}
\usepackage{footmisc}
\usepackage{soul}

\usepackage{natbib,twoopt}
\bibpunct{(}{)}{;}{a}{}{,} 
\makeatletter
\newcommandtwoopt{\citeads}[3][][]{\href{http://adsabs.harvard.edu/abs/#3}%
{\def\hyper@linkstart##1##2{}%
\let\hyper@linkend\@empty\citealp[#1][#2]{#3}}}
\newcommandtwoopt{\citepads}[3][][]{\href{http://adsabs.harvard.edu/abs/#3}%
{\def\hyper@linkstart##1##2{}%
\let\hyper@linkend\@empty\citep[#1][#2]{#3}}}
\newcommandtwoopt{\citetads}[3][][]{\href{http://adsabs.harvard.edu/abs/#3}%
{\def\hyper@linkstart##1##2{}%
\let\hyper@linkend\@empty\citet[#1][#2]{#3}}}
\newcommandtwoopt{\citeyearads}[3][][]%
{\href{http://adsabs.harvard.edu/abs/#3}
{\def\hyper@linkstart##1##2{}%
\let\hyper@linkend\@empty\citeyear[#1][#2]{#3}}}
\makeatother

\bibpunct{(}{)}{;}{a}{}{,}

\bibpunct{(}{)}{;}{a}{}{,} 

\newcommand \sw{{\it Swift}}
\newcommand \fe{{\it Fermi}}
\newcommand \eb{$E_{\rm break}$}
\newcommand \ep{$E_{\rm peak}$}
\newcommand \gc{$\gamma_{\rm c}$}
\newcommand \gm{$\gamma_{\rm m}$}
\newcommand \ec{$E_{\rm c}$}
\newcommand \fnu{$F_\nu$}
\newcommand \nufnu{$\nu F_\nu$}
\newcommand \rat{$\gamma_{\rm m}/\gamma_{\rm c}$}
\newcommand{\order}[1]{} 

\def\gsim{ \lower .75ex \hbox{$\sim$} \llap{\raise .27ex \hbox{$>$}} }

\def\lsim{ \lower .75ex\hbox{$\sim$} \llap{\raise .27ex \hbox{$<$}} }

\begin{document}

\title{Prompt optical emission as a signature of synchrotron radiation in gamma-ray bursts}

\author{G. Oganesyan\inst{\ref{inst1},\ref{inst2},\ref{inst3}}
\and L. Nava\inst{\ref{inst4},\ref{inst5},\ref{inst6}}
\and G. Ghirlanda\inst{\ref{inst4},\ref{inst7}}
\and A. Melandri\inst{\ref{inst4}}
\and A. Celotti\inst{\ref{inst1},\ref{inst4},\ref{inst6}}
}
\institute{SISSA, via Bonomea 265, I--34136 Trieste (TS), Italy
\label{inst1}
\and
Gran Sasso Science Institute, Viale F. Crispi, 7, I--67100 L'Aquila (AQ), Italy
\email{\href{mailto:gor.oganesyan@gssi.it}{gor.oganesyan@gssi.it}}
\label{inst2}
\and 
INFN -- Laboratori Nazionali del Gran Sasso, I--67100, L'Aquila (AQ), Italy
\label{inst3}
\and
INAF -- Osservatorio Astronomico di Brera, via E. Bianchi 46, I--23807 Merate (LC), Italy \label{inst4}
\and
INAF -- Osservatorio Astronomico di Trieste, via G.B. Tiepolo 11, I--34143 Trieste, Italy \label{inst5}
\and
INFN -- Istituto Nazionale di Fisica Nucleare, Sezione di Trieste, via Valerio 2, I--34127, Trieste, Italy \label{inst6}
\and
INFN -- Milano Bicocca, Piazza della Scienza 3, I--20123, Milano, Italy \label{inst7}
}
\abstract{
Information on the spectral shape of prompt emission in gamma-ray bursts (GRB) is mostly available only at energies $\gtrsim10$\,keV, where the main instruments for GRB detection are sensitive. The origin of this emission is still very uncertain because of the apparent inconsistency with synchrotron radiation, which is the most obvious candidate, and the resulting need for considering less straightforward scenarios.
The inclusion of data down to soft X-rays ($\sim$\,0.5\,keV), which are available only in a small fraction of GRBs, has firmly established the common presence of a spectral break in the low-energy part of prompt spectra, and even more importantly, the consistency of the overall spectral shape with synchrotron radiation in the moderately fast-cooling regime, the low-energy break being identified with the cooling frequency. 
In this work we further extend the range of investigation down to the optical band. In particular, we test the synchrotron interpretation by directly fitting a theoretically derived synchrotron spectrum and making use of optical to gamma-ray data. Secondly, we test an alternative model that considers the presence of a black-body component at $\sim$keV energies, in addition to a non-thermal component that is responsible for the emission at the spectral peak (100\,keV--1\,MeV). 
We find that synchrotron radiation provides a good description of the broadband data, while models composed of a thermal and a non-thermal component require the introduction of a low-energy break in the non-thermal component in order to be consistent with optical observations.
Motivated by the good quality of the synchrotron fits, we explore the physical parameter space of the emitting region. In a basic prompt emission scenario we find quite contrived
solutions for the magnetic field strength (5\,G~$<B^\prime<40$\,G) and for the location of the region where the radiation is produced ($R_\gamma>10^{16}$\,cm). We discuss which assumptions of the basic model would need to be relaxed in order to achieve a more natural parameter space.
}


\maketitle

\section{Introduction}\label{sec:intro}

Among the many unsolved questions that still limit our comprehension of the physics involved in the gamma-ray burst (GRB) phenomenon, the origin of the prompt radiation is one of the most fundamental.
The difficulty of understanding the prompt emission phase encompasses the whole process, from the nature of the energy that powers the emission to the specific radiative process that is responsible for the radiation.
The standard approach that is adopted to investigate the origin of prompt emission involves fitting empirical functions to prompt spectra and comparing the values of the best-fit parameters with expectations from different radiative processes and in particular from synchrotron radiation, the most natural candidate.
Two smoothly connected power laws (PLs) usually provide reasonable fits.
This type of studies has revealed a significant inconsistency between observations and the synchrotron mechanism, the low-energy photon index being on average harder ($\langle\alpha\rangle\simeq-1$) than the fast-cooling synchrotron value ($\langle\alpha^{\rm syn}\rangle=-3/2$)  \citep{Preece98,ghisellini00,Frontera00,Ghirlanda02,Kaneko06,Nava11}. 

A recent breakthrough discovery came from the investigation of prompt emission spectra in the soft X-ray band in cases where the X-ray Telescope (XRT, 0.3-10\,keV) on board the {\it Neil Gehrels Swift Observatory} (\sw\ hereafter) was able to observe the prompt emission phase, thus extending the spectral information well below 10\,keV.
By performing a time-resolved joint spectral analysis of simultaneous \sw/XRT+BAT data on 14 long GRBs, \cite{Oganesyan2017} showed that more than 60\% of the spectra require a fitting function with an additional harder PL segment to describe the spectrum below $2-20$\,keV. In other words, a good description of the data requires a spectral break at energies $\lesssim20$\,keV.
These results have later been confirmed by a similar investigation conducted on a larger sample of 34 XRT+BAT events \citep{gor2018}. 
The same shape was also found in GRB~160625B, one of the brightest \fe-GBM GRBs \citep{ravasio2018}.
Moreover, a recent investigation has pointed out that these types of spectral breaks are a recurring feature in bright GBM bursts and can extend to energies around $\sim$\,100\,keV \citep{Ravasio2019}. 
All these studies point to similar results, with most of the spectra requiring a fitting function featuring three PL segments: i) a hard PL at low energies (up to a few keV or tens of keV) with an average photon index $\langle\alpha_1\rangle\simeq-2/3$, ii) a second softer PL segment with an average photon index $\langle\alpha_2\rangle\simeq-3/2$ describing the spectrum at intermediate energies, and iii) a third PL with a photon index $\beta<-2$ at higher energies. These three PLs identify two break energies: \eb, typically located between 1-20 keV but also extending to 100\,keV in bright GBM bursts, and \ep, which corresponds to the peak of the \nufnu\ spectrum, and is usually located between 0.1 and 1\,MeV.
Remarkably, the typical values of all photon indices are in good agreement with the values that are predicted when the dominant emission mechanism is synchrotron radiation.

A comparison based only on photon indices, however, is not sufficient to claim a synchrotron origin. 
In particular, two main critiques that are usually raised against the synchrotron interpretation must be addressed: the first concerns the spectral width that might be too narrow in the observed spectra as compared to synchrotron spectra (\citealt{Beloborodov2013,Axelsson2015,Yu2015,Vurm2016}, see however \citealt{Burgess2017}), and the second concern is based on the observation that in a small but sizable fraction of GRBs the low-energy photon index is harder than the limiting synchrotron value -2/3 (the so-called line of death, \citealt{Preece98,Kaneko06}).
Moreover, an alternative interpretation to the low-energy spectral break has been proposed, and invokes the presence of a sub-dominant black-body (BB) component (in addition to a non-thermal component) that peaks at $\sim$\,10-50\,keV \citep{guiriec2011,guiriec2013,ghirlanda2013,axelsson2012,iyyani2013,peng2014}. 
While it is now clear that a simple double PL is not sufficient to capture the shape of prompt spectra, it is less clear which of the two different proposed models is correct: the model that at low energies adds a third PL segment (and then describes the whole spectrum with one single component), or the model that invokes an additional (thermal) spectral component. 
The question is of paramount importance because the two different interpretations of the spectral shape imply two very different theoretical scenarios, with different implications on the nature of the jet energy, and possibly on the location of the dissipation region and nature of the dissipation mechanism.

Observations extending to lower frequencies ($\ll$\,0.1\,keV) would allow us to address all these questions by determining in a more robust way i) the low-energy spectral index (testing if its value crosses the line of death), ii) the spectral width (through the use of a synchrotron fitting function in place of empirical functions), and iii) distinguishing among the two competing spectral models (that predict very different optical fluxes when they are extrapolated to the optical band).  
An important tool for testing prompt emission spectral models is then the inclusion of optical observations simultaneous to X/$\gamma$-ray observations. 
Prompt optical observations were successfully performed for a limited number of GRBs with the Ultra-Violet/Optical Telescope (UVOT, \citealt{uvot}) on board {\it Swift} and with ground-based robotic telescopes such as Robotic Optical Transient Search Experiment (ROTSE-III, \citealt{Akerlof2003}), Rapid Action Telescope for Transient Objects (TAROT, \citealt{Klotz2009}), the Mobile Astronomical System of Telescope-Robots (MASTER, \citealt{Lipunov2004}), Pi of the Sky \citep{Burd2005}, and Telescopio Ottimizzato per la Ricerca dei Transienti Ottici RApidi (TORTORA, \citealt{Beskin2017}). 

One caveat for the use of optical observations as a test for prompt emission models is, however, a possible contamination from emission of a different origin, for example, forward and/or reverse shock radiation generated by the deceleration of the GRB outflow by the external medium.
A first indication of the internal or external origin of the early-time optical emission comes from its temporal behaviour, which is expected to track the variability that is detected in the hard X-ray band when the two share a common internal origin.
These different behaviours have indeed been pointed out in the study of prompt optical emission, and have shown that its origin is not always the same and can vary for each case (see \citealt{Kopac2013} for a recent systematic investigation).

In this paper we address the problem of the consistency of prompt spectra with synchrotron radiation and of the validity of the synchrotron versus BB+non-thermal modelling by means of i) the inclusion of optical data and ii) the use of a synchrotron model (in place of empirical functions) for spectral fitting.
We use a sample of 21~GRBs that are characterised by simultaneous optical, X--ray (0.3-10 keV), and $\gamma$--ray (15-150 keV) prompt observations. 
Among these cases, we focus on those for which the optical emission most likely has an internal origin, that is, the variability in the optical band tracks the variability in the soft and hard X-ray bands.
We also analyse those cases where the temporal behaviour of the optical emission suggests a different origin for this emission, and cases where no temporal information is available (i.e. only a single-epoch optical observation is available during the prompt emission). 
Interesting conclusions can also be deduced from the analysis of these cases.
Finally, motivated by the success in fitting a synchrotron spectrum to the data, we calculate the values of the Lorentz factor, magnetic field, distance of the emitting region, and typical electron Lorentz factor required in the emitting region to explain the observations, assuming a standard scenario. We show that the inferred values are quite contrived and most likely demand a reconsideration of the standard model.

\section{Sample and data extraction}\label{sect:method}
We started from the sample collected by \cite{gor2018}, composed of 34~GRBs with XRT observations of the prompt emission phase. 
For each burst we collected from the literature all available optical observations (detections and upper limits) that lay within the same temporal window as defined by simultaneous XRT+BAT observations.
When available, we collected published calibrated magnitudes, otherwise we considered the information reported in the GRB Circular Network (GCNs). 
We corrected the observed magnitudes for Galactic extinction (according to 
\citealt{schlafly11}) and for extinction in the host galaxy (if known). 
Lastly, we converted the de-reddened magnitudes into flux densities. 
The requirement of having at least one optical observation during the prompt emission phase limits the final sample to 21~GRBs.
Table~\ref{tab:table_optical} lists for each GRB the time intervals of optical observations, flux density, filter, the information whether the correction for extinction was estimated only for our Galaxy (G) or also for the host galaxy (HG), and the reference.

For each time interval defined by the epochs of optical observations, we analysed the X-ray and $\gamma$-ray spectra following the same procedure as was adopted in \cite{Oganesyan2017}. Here we summarise the main steps of the procedure we followed to extract the data from the different instruments. 

BAT event files were downloaded from the \sw\ data archive\footnote{\url{http://heasarc.gsfc.nasa.gov/cgi-bin/W3Browse/swift.pl}}. 
We extracted BAT spectra and light curves with the latest version of the {\sc heasoft} package (v6.17). We used the {\tt FTOOLS} {\tt batmaskwtevt} and {\tt batbinevt} to extract the background-subtracted mask-weighted BAT light curves in the energy range 15-150\,keV. BAT spectral files were generated using the {\tt batbinevt} task. The spectral files were corrected with {\tt batupdatephakw} and {\tt batphasyserr} to include systematic errors. 
We generated response matrices for time intervals before, during, and after the satellite slew using the {\tt batdrmgen} tool. 
The latest calibration files (CALDB release 2017-05-20) were adopted.

XRT light curves were downloaded from the \sw\ Science Data Center, provided by the University of Leicester\footnote{\url{http://www.swift.ac.uk/xrt_curves/}} \citep{Evans_09}.  
XRT event files were retrieved from the \sw/XRT archive\footnote{\url{http://www.swift.ac.uk/archive/}}. We extracted source and background spectra in each time-bin using  the {\tt xselect} tool.
To avoid pile-up effects, we removed the central region of the XRT images, following the procedure suggested in \cite{Romano_06}. Ancillary response files were generated using the task {\tt xrtmkarf}. 
For the spectral analysis, we excluded all channels below 0.5\,keV.
To use $\chi^2$ statistics, energy channels were grouped using the {\it grppha} tool and requiring at least 20 counts per bin.  

\fe/GBM observations are available only for four~GRBs and were included in the analysis. 
We considered the data from the two most illuminated NaI (in the range 8-1000\,keV) and from one BGO detector (300\,keV-40\,MeV). 
We excluded channels in the range 30-40\,keV due to the presence of the iodine K-edge at 33.17\,keV. We extracted the spectra using the \textsc{gtburst} tool\footnote{\url{http://fermi.gsfc.nasa.gov/}}. 
To use $\chi^2$ statistics, energy channels were grouped using the {\it grppha} tool and requiring at least 20 counts per bin.

\section{Spectral analysis}\label{sec:analysis}
We performed a joint spectral analysis of XRT+BAT (or XRT+BAT+GBM) data with two different models: a synchrotron model, and a two-component model (more specifically, a BB plus a PL with a high-energy cutoff, called hereafter BB+CPL model). 
Spectral analysis (performed with XSPEC v12.9.1) was limited to temporal bins where optical observations are available (see Table~\ref{tab:table_optical}). 
Optical data were not included in the spectral fitting. 
Their consistency with the spectral shape defined by spectral data at energies $>$\,0.5\,keV was tested by extrapolating the best-fit models down to the optical band. 
The results of this analysis are presented in  Sect.\ref{sec:results}.

We took into account both Galactic and intrinsic absorption of X-ray spectra by neutral hydrogen using the multiplicative XSPEC models {\tt tbabs} and {\tt ztbabs} \citep{wilms00}. 
The Galactic column density of neutral hydrogen in the direction of a GRB is estimated from \citet{Kalberia_05}. 
The intrinsic column densities were taken from \citet{gor2018}, where the values have been derived from the analysis of the late-time X-ray spectrum during the afterglow phase. The procedure is described in detail in \citet{Oganesyan2017}. 

In order to account for the uncertainty in the inter-calibration between the XRT and BAT, we allowed for a 10\% discrepancy in the normalisation factor of one of the two instruments. Because we aim to extrapolate the best-fit model down to the optical band and compare it with the measured optical flux, slightly different results might be derived depending on whether we decide to fix the XRT normalisation and allow for uncertainties in the BAT normalisation. 
We then fixed first the normalisation of XRT and allowed for a 10\% variation in the normalisation of BAT, and then repeated the spectral analysis by fixing BAT and allowing for a 10\% variation in XRT. 
When GBM observations were also available, we included them in the joint spectral fitting, considering a possible 10\% uncertainty in the calibration of the GBM as compared to \sw\ instruments.

\subsection{Synchrotron fits}
No synchrotron model is available in XSPEC, therefore we added the possibility to fit synchrotron spectra as a \textup{{\it table model}}. 
We considered a population of electrons accelerated into a  PL energy distribution $dN_{\rm e}/d\gamma \propto \gamma^{-p}$ for Lorentz factors $\gamma$ between a minimum value \gm\ and a maximum value $\gamma_{\rm max}\gg\gamma_{\rm m}$.
We considered electron cooling through synchrotron and inverse Compton radiation in the Thomson regime.
In case of the fast-cooling regime (i.e. for $\gamma_{\rm c}<\gamma_{\rm m}$), after a time $t$ the average distribution is $dN_{\rm e}/d\gamma \propto \gamma^{-2}$ for $\gamma_{\rm c}<\gamma<\gamma_{\rm m}$ and $dN_{\rm e}/d\gamma \propto \gamma^{-p-1}$ for $\gamma>\gamma_{\rm m}$, where $\gamma_{\rm c}$ is the cooling Lorentz factor.
In the slow-cooling regime, the distribution is approximated by $dN_{\rm e}/d\gamma \propto \gamma^{-p}$ for $\gamma_{m}<\gamma<\gamma_{c}$  and $dN_{\rm e}/d\gamma \propto \gamma^{-p-1}$ for $\gamma>\gamma_{\rm c}$. 

The photon spectrum was calculated by integrating the single-electron spectrum over the electron distribution.
The overall {\it \textup{shape}} of the photon spectrum depends only on two quantities: the ratio \rat\ , and $p$. 
We generated synchrotron spectra for values of \rat\ ranging from 0.1 to 100 (embracing both the slow- and fast-cooling regimes), and for values of $p$ between 2 and 5. 
The synchrotron model was implemented in XSPEC in the form of a tabulated additive model. 
The final model has then four free parameters: \rat, $p$, \ec, and normalisation. 
Here \ec\,$=h\,\nu_{\rm c}$ is the energy corresponding to the cooling frequency $\nu_{\rm c}$, that is, the synchrotron frequency of electrons with Lorentz factor \gc.

We find that the high-energy part of the spectrum is almost never constrained because GBM data are missing in most cases. 
The value of $p$  is constrained only in one GRB (for which GBM data are indeed available).
For all the other cases, we fixed the value of $p$ to 2.6, motivated by the typical value of the high-energy photon index $\beta\sim-2.3$ found from spectral analysis of large samples with empirical models \citep[e.g.][]{Kaneko06,gruber14,goldstein12,nava2011}. 
Within a synchrotron interpretation, the values of $p$ and $\beta$ are indeed related by $\beta=-\frac{p+2}{2}$.

\begin{figure}
\begin{center}
   {\centering
  \includegraphics[width=0.48\textwidth]{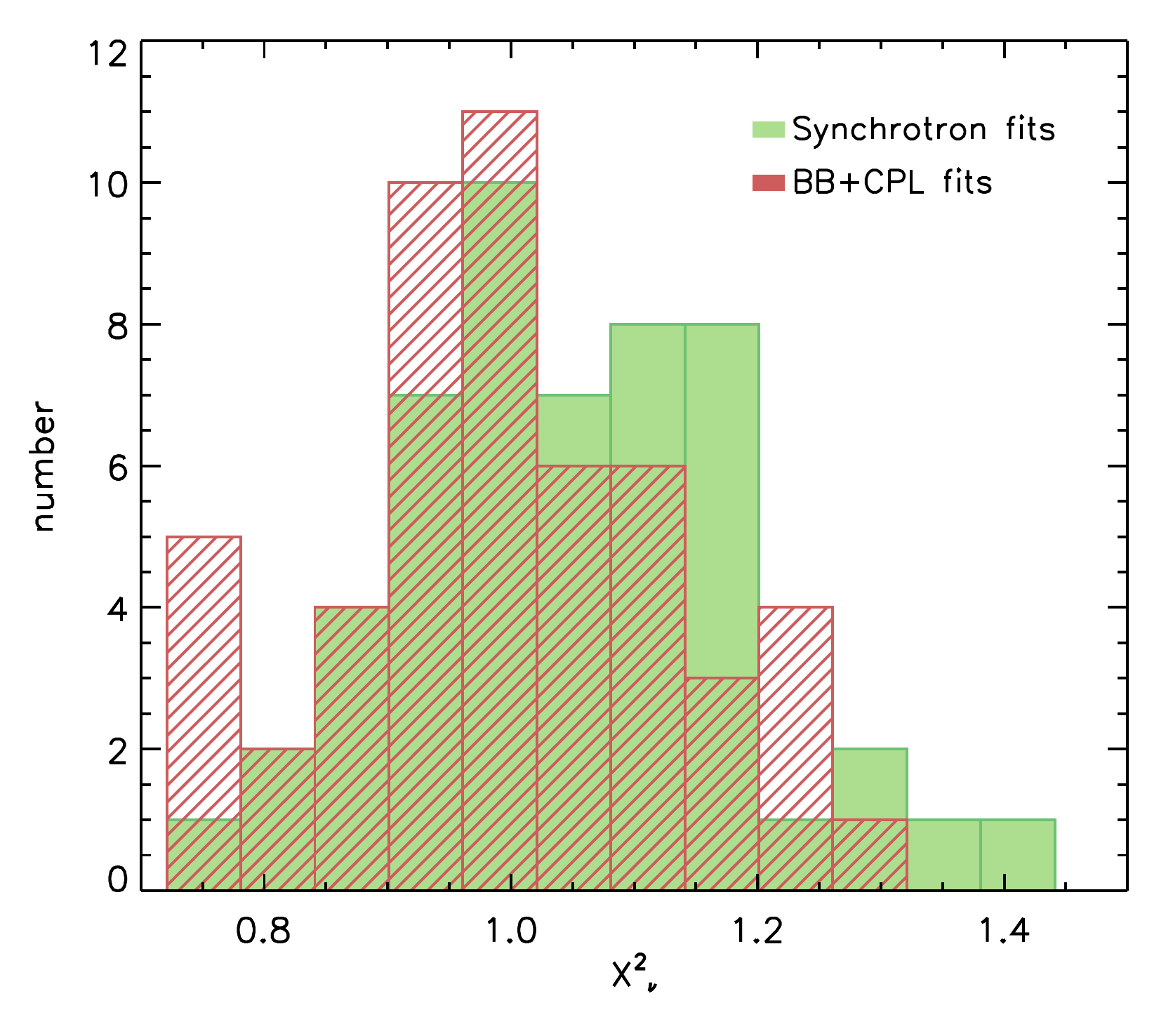}
  }
\caption{\label{fig:chisq} Distribution of the reduced chi-square $\chi^2_{\nu}$ for synchrotron (green histogram) and BB+CPL (red hatched histogram) fits. All 52 time-resolved spectra (from 21 GRBs) are included.
}
\end{center}
\end{figure}

\begin{figure*}
\begin{center}
   {\centering
  \includegraphics[width=0.48\textwidth]{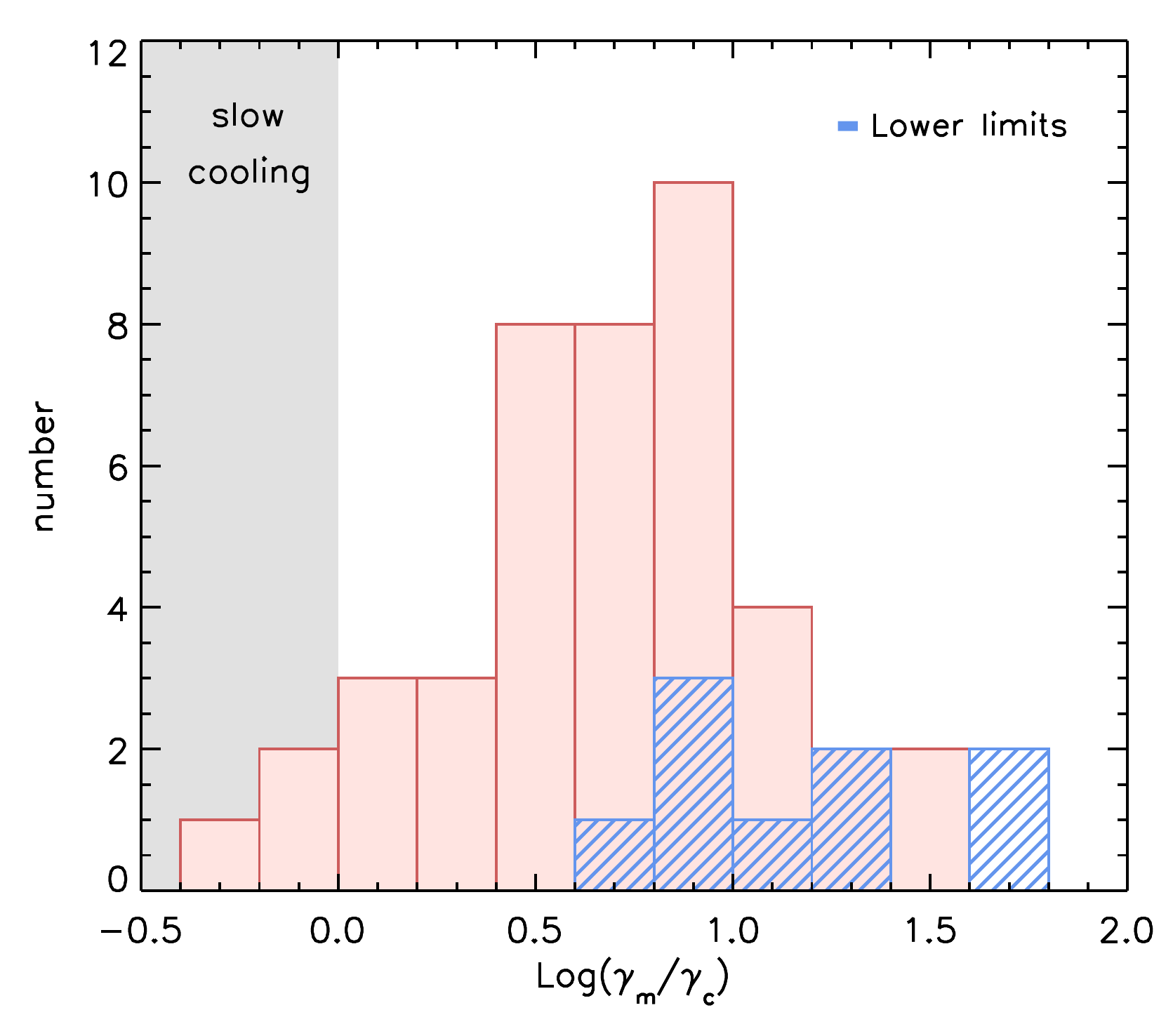}
  \hskip 0.46truecm
  \includegraphics[width=0.48\textwidth]{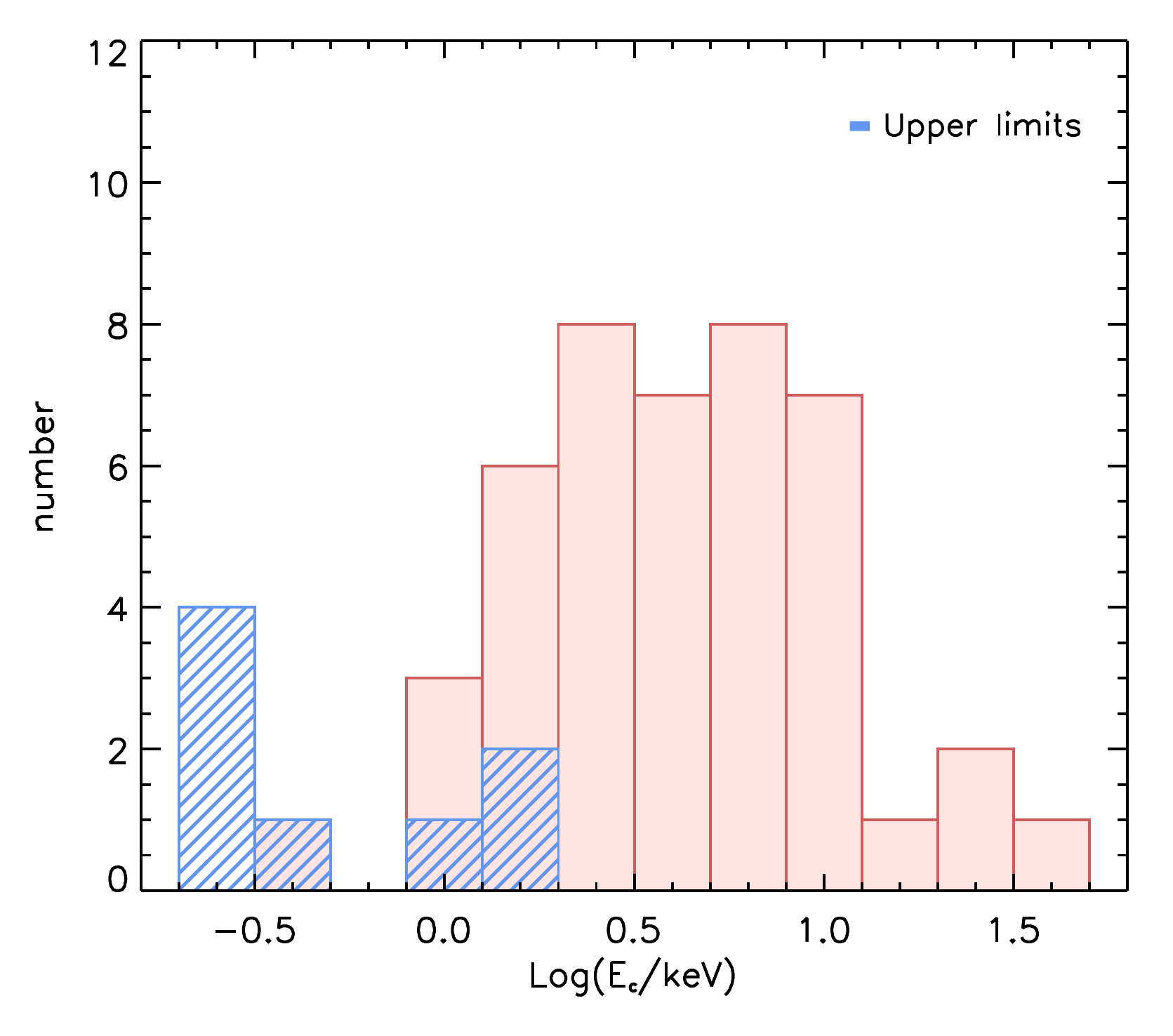}
  }
\caption{\label{fig:syn_results} Distributions of the best-fit values for the parameters of the synchrotron model for the full sample of 52 time-resolved spectra (21 GRBs). Left panel: distribution of the ratio between \gm\ and \gc. The grey shaded area highlights the region corresponding to the slow-cooling regime (\gm\,$<$\,\gc). Right panel: Distribution of the cooling energy \ec.}
\end{center}
\end{figure*}
\begin{figure}
  \includegraphics[width=0.48\textwidth]{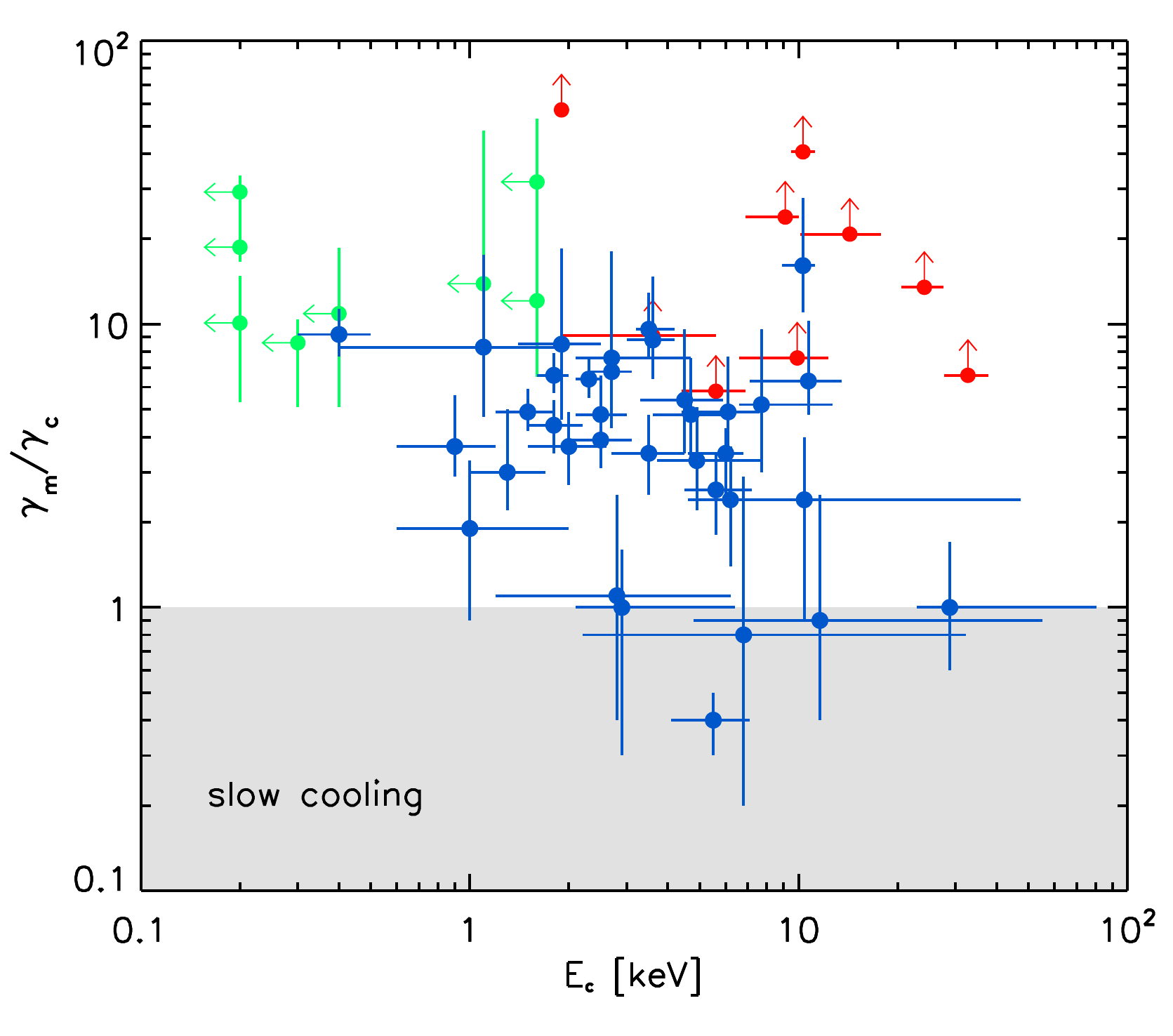}
  \caption{Best-fit values for the parameters of the synchrotron model for the full sample of 52 time-resolved spectra (21 GRBs). The ratio \rat\ is shown as a function of the cooling energy \ec. The grey shaded area highlights the region corresponding to the slow-cooling regime (\gm\,$<$\,\gc). Leftwards-pointing arrows correspond to cases where the \ec\ value is an upper limit. Upwards-pointing arrows correspond to cases where only a lower limit can be placed on the ratio \rat. }
  \end{figure}

The results of synchrotron fits are reported in Table~\ref{tab:table_fits} and shown in Figs.~\ref{fig:061121}-\ref{fig:110205A}
for three peculiar cases (see below) and in Appendix~\ref{app:multi} and \ref{app:single} for the rest of the sample. 
For each GRB and for each time bin, Table~\ref{tab:table_fits} reports the synchrotron cooling energy \ec, the ratio between the characteristic electron Lorentz factors \rat, the density flux $F_{\rm c}$ at \ec, the $\chi^{2}$ , and the degrees of freedom (d.o.f.). 
The synchrotron model provides acceptable fits, with $\chi^2_\nu<1.2$ (with the exception of 5 cases where $1.2<\chi^2_\nu<1.4$), and associated null hypothesis probabilities higher than $10^{-2}$ for 51 spectra out of 52. 
The distribution of the reduced $\chi^2_\nu$ is shown in Fig.~\ref{fig:chisq} (green histogram).
No issue is found with the spectral width around the peak energy, which is well described by the synchrotron spectral model.

The distributions of the model parameters \rat\ and \ec\ are shown in Fig.~\ref{fig:syn_results}. 
The ratio \rat\ (left panel) is in the range 0.3--30, with a few cases corresponding to the slow-cooling regime (i.e. $\gamma_{\rm m}<\gamma_{\rm c}$, grey shaded area).
Assuming a log-normal distribution, the mean value is $\langle\,\rm{Log}(\gamma_{\rm m}/\gamma_{\rm c})\rangle=0.56$ (and the dispersion $\sigma=0.36$), corresponding to a typical value \rat$\,\sim 4$. 
The \ec\ distribution (right panel) is described by a log-normal function with a mean value $\langle$Log$(E_{\rm c}/\rm keV)\rangle=0.53$ ($\sigma = 0.37$), corresponding to \ec\,$\sim$\,3\,keV. The value of $p$ is constrained only for one spectrum (GRB 100906A), and its best-fit value is very steep: $p=4.4_{-0.4}^{+0.5}$.

An actual synchrotron spectrum has rarely been used to fit prompt emission spectra. 
Few {\it BATSE} GRB spectra have been modelled with a synchrotron spectrum in the slow-cooling regime \citep{tavani1996} and for high self-absorption frequencies \citep{Lloyd00}.
More recently, the synchrotron radiation spectrum has been found to successfully fit the time-resolved spectra of GRB\,130606B \citep{zhang2016} and GRB\,160625B \citep{zhang2018}. 
A recent investigation of 19 bright single-pulse {\it Fermi} GRBs revealed that most of the time-resolved spectra can be successfully fitted by the synchrotron model when cooling of the electrons is taken into account \citep{burgess2018}. 
Our analysis is in agreement with the findings described above: in the considered sample, the synchrotron model can account for the prompt emission spectra if the electron cooling is not complete. 

\subsection{BB+CPL fits}
The results of CPL+BB fits are reported in Table~\ref{tab:table_fits} and shown in  Figs.~\ref{fig:061121}-\ref{fig:110205A}
for three peculiar cases (see below) and in Appendix~\ref{app:multi} and \ref{app:single} for the rest of the sample. For each GRB and for each time bin, Table~\ref{tab:table_fits} reports the low-energy photon index $\alpha$ and the peak energy $E_{\rm p}$ of the CPL, the temperature of the BB component $kT$, the $\chi^{2}$ , and the degrees of freedom (d.o.f.). 
Exactly as for the synchrotron model, the BB+CPL model also provides acceptable fits with reduced chi-square $\chi^2_\nu<1.2$ with the exception of five cases. 
The associated null hypothesis probabilities are higher than $10^{-2}$ for all 52 spectra. 
The distribution of the reduced $\chi^2_\nu$ is shown in Fig.~\ref{fig:chisq} (red hatched histogram).

The low-energy photon index $\alpha$ of the CPL component is in the range between -0.80 and -1.80 with a mean value of $\langle\alpha\rangle=-1.25$ and dispersion $\sigma=0.24$ (a normal distribution has been assumed). The log-normal distribution of the peak energy returns a mean value $\langle$Log$(E_{\rm p}/\rm keV)\rangle=1.86$ ($\sigma = 0.36$), which corresponds to a typical value $E_{\rm p}$\,$\sim$\,72\,keV. The temperature of the BB component is described by a log-normal distribution with a mean value $\langle$Log$(kT/\rm keV)\rangle=0.18$ ($\sigma = 0.27$), corresponding to a typical $kT$\,$\sim $\,1.50\,keV. In two spectra no BB component is required by the data.\\

The results for the spectral fits that were obtained with the two different models outline the difficulty of distinguishing between the two scenarios based on data in the 10-10$^3$\,keV energy range alone. In the next section we introduce optical observations.


\section{Testing models with prompt optical observations}\label{sec:results}
\begin{figure*}
\begin{center}
\includegraphics[scale=0.67]{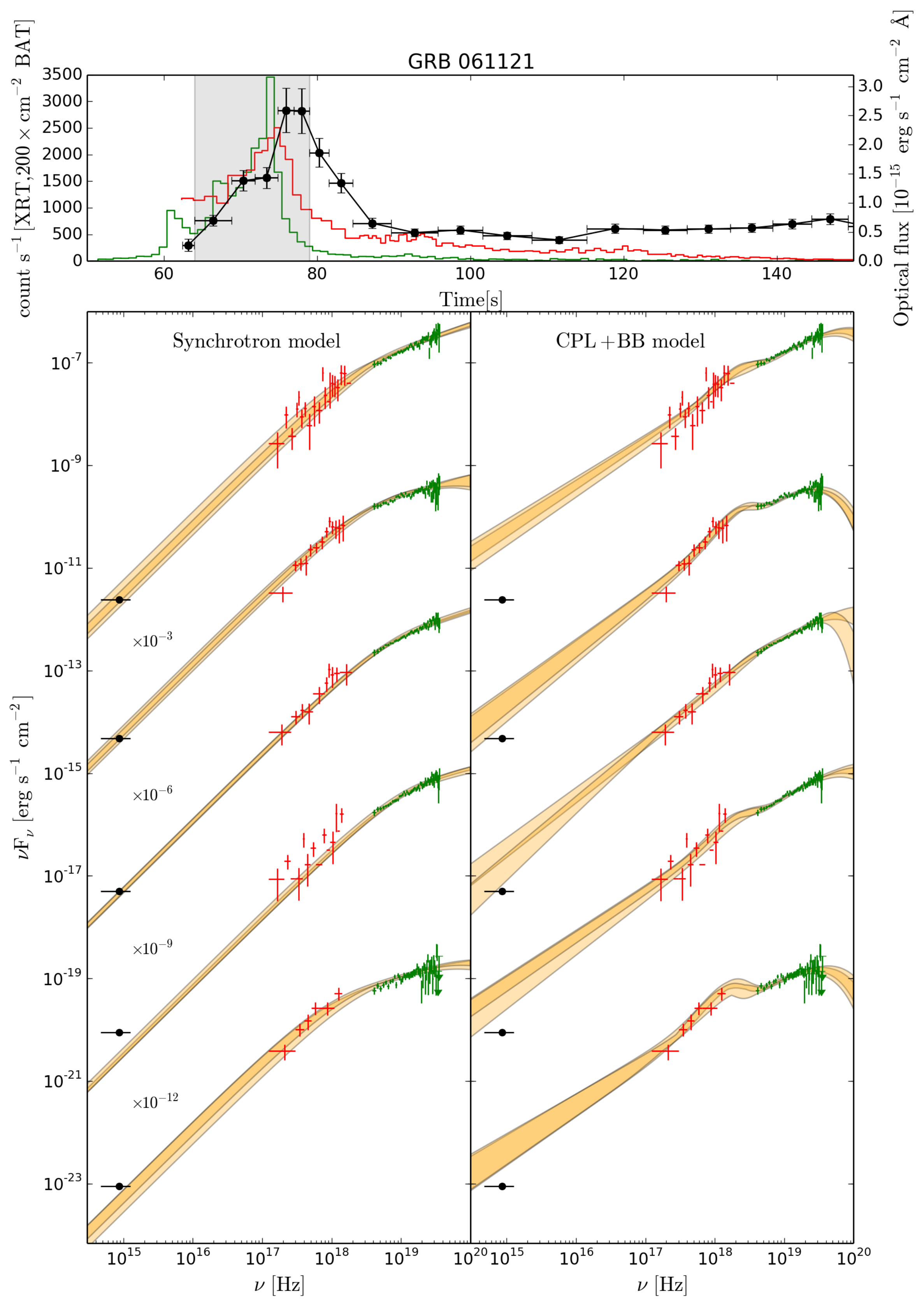}
\caption{Results of the time-resolved spectral analysis of the prompt emission of GRB\,061121. Upper panel: XRT (red), BAT (green), and optical (black) light curves. Bottom panels: Synchrotron (left) and BB+CPL (right) fits to XRT (red) and BAT (green) spectral data corresponding to the time interval highlighted in grey in the upper panel. Black filled circles show the optical flux. XRT spectra are de-absorbed. The best-fit confidence regions are shown in orange: light orange for contours derived when the normalisation of BAT data is kept fixed and that of XRT is free to vary between 0.9 and 1.1, and dark orange contours for the opposite situation. }
\label{fig:061121} 
\end{center}
\end{figure*}
\begin{figure*}
\begin{center}
\ContinuedFloat
\includegraphics[scale=0.67]{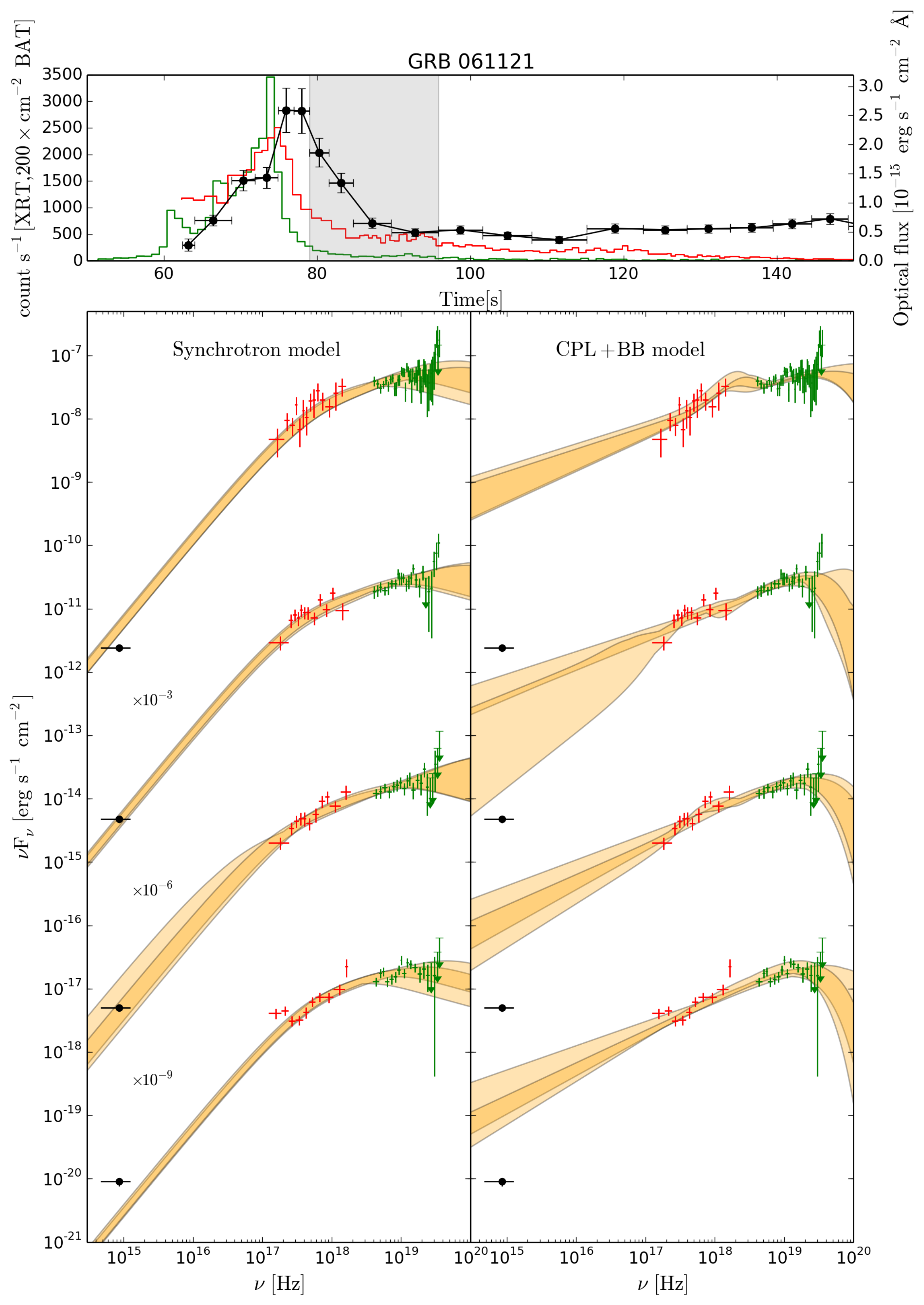}
\caption{Continuation.}
\end{center}
\end{figure*}
\begin{figure*}
\begin{center}
\includegraphics[scale=0.67]{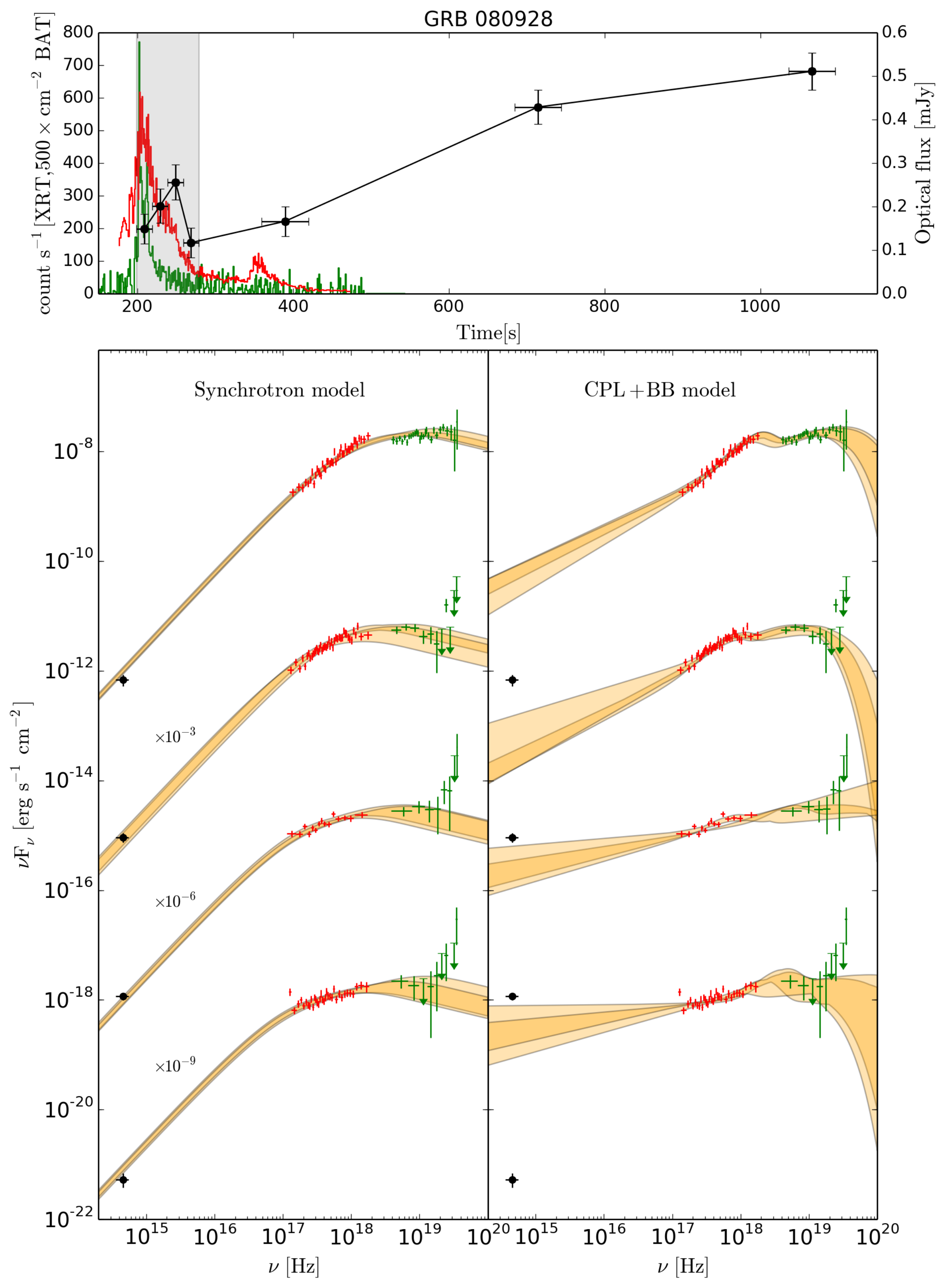}
\caption{Results of time-resolved spectral analysis of the prompt emission of GRB\,080928. Upper panel: XRT (red), BAT (green), and optical (black) light curves. Bottom panels: Synchrotron (left) and BB+CPL (right) fits to XRT (red) and BAT (green) spectral data  corresponding to the time interval highlighted in grey in the upper panel. Black filled circles show the optical flux. XRT spectra are de-absorbed. The best-fit confidence regions are shown in orange: light orange for contours derived when the normalisation of BAT data is kept fixed and that of XRT is free to vary between 0.9 and 1.1, and dark orange contours for the opposite situation. }
\label{fig:080928} 
\end{center}
\end{figure*}
\begin{figure*}
\begin{center}
\includegraphics[scale=0.67]{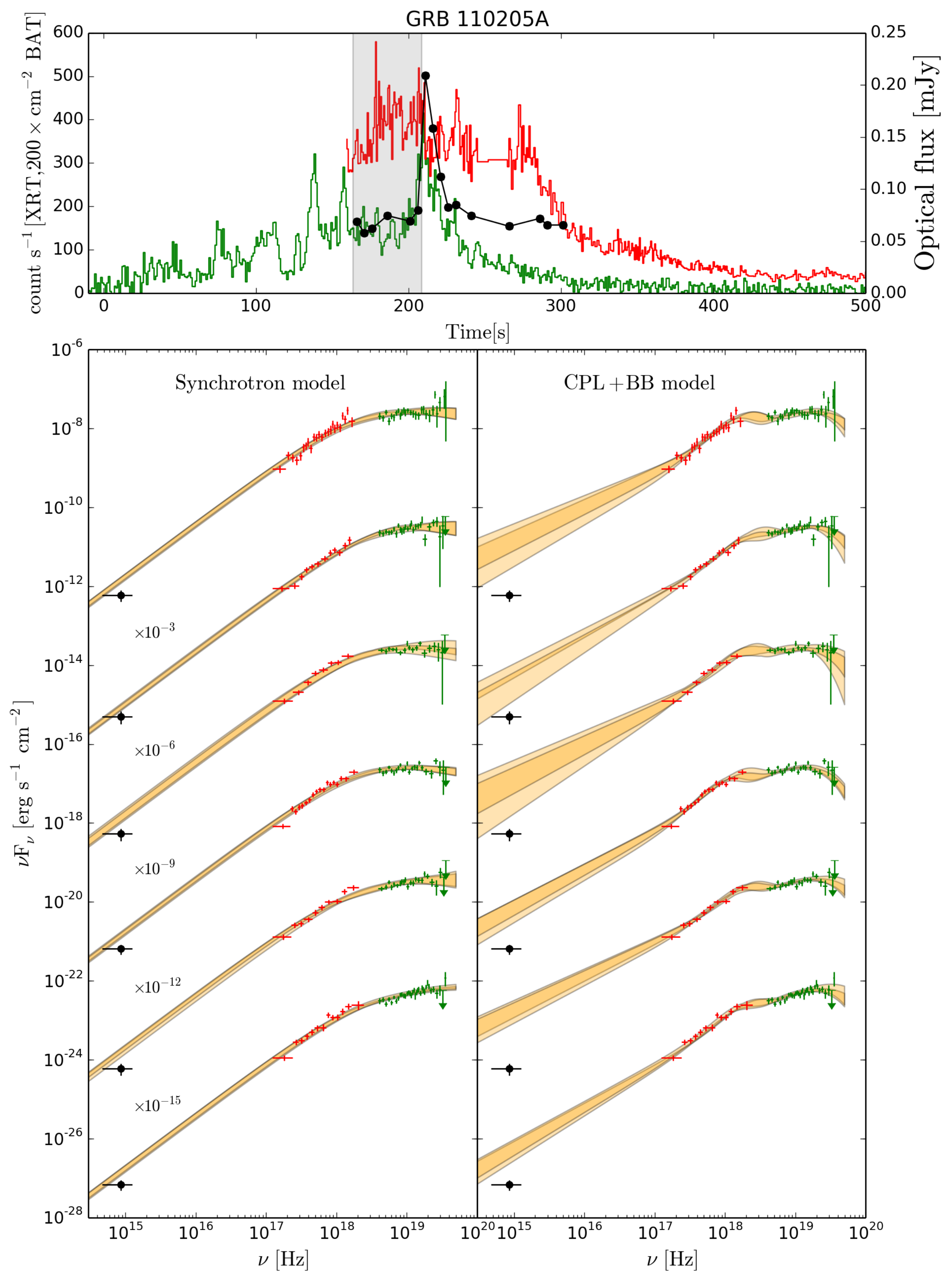}
\caption{\label{fig:110205A} Results of time-resolved spectral analysis of the prompt emission of GRB\,110205A. Upper panel: XRT (red), BAT (green), and optical (black) light curves. Bottom panels: Synchrotron (left) and BB+CPL (right) fits to XRT (red) and BAT (green) spectral data corresponding to the time interval highlighted in grey in the upper panel. Black filled circles show the optical flux. XRT spectra are de-absorbed. The best-fit confidence regions are shown in orange: light orange for contours derived when the normalisation of BAT data is kept fixed and that of XRT is free to vary between 0.9 and 1.1, and dark orange contours for the opposite situation.}
\end{center}
\end{figure*}
\begin{figure*}
\begin{center}
\ContinuedFloat
\includegraphics[scale=0.64]{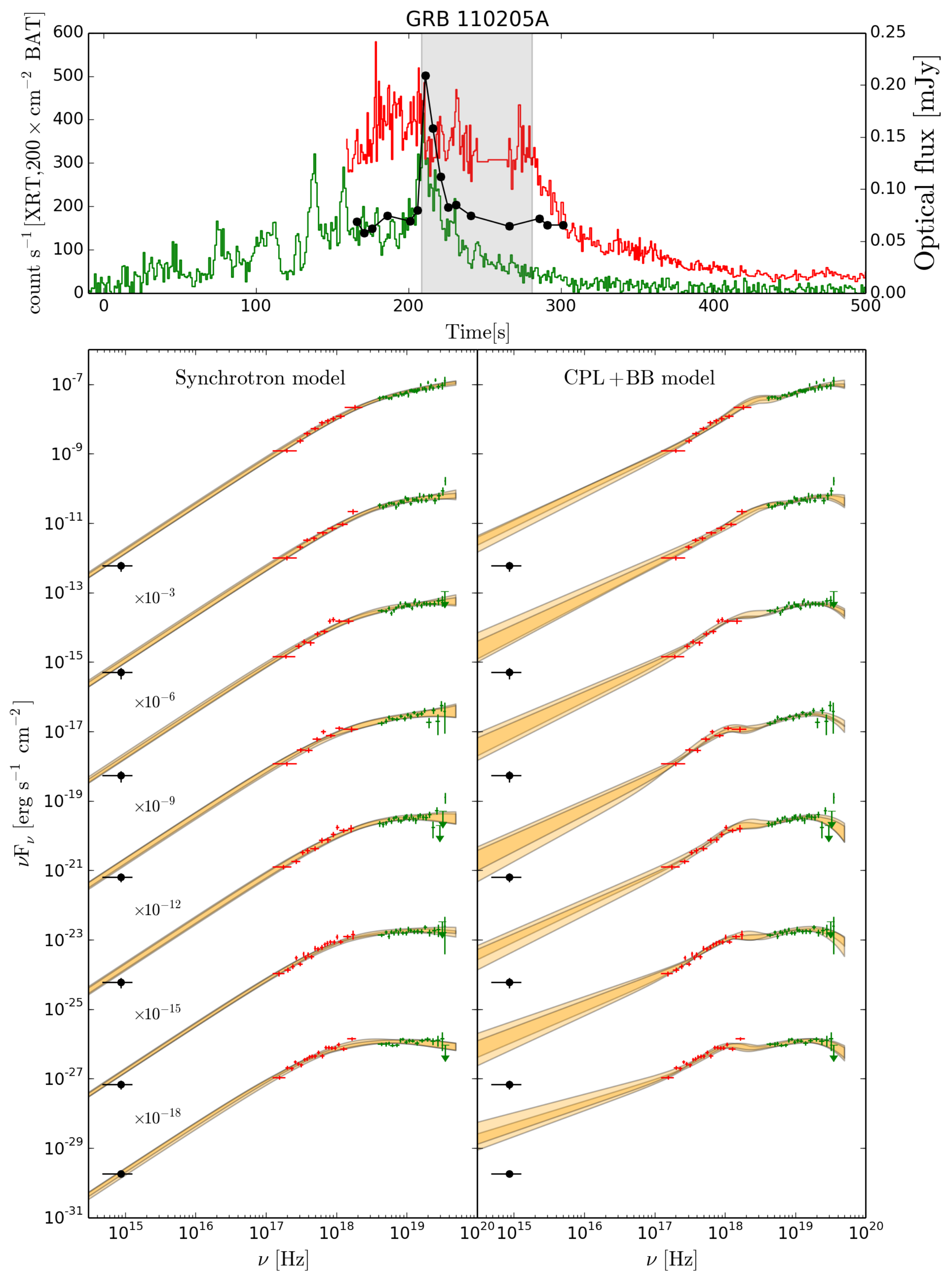}
\caption{Continuation.}
\end{center}
\end{figure*}

The full sample of 21 GRBs contains 3 cases with multi-epoch optical observations during the prompt emission and optical variability that tracks the XRT and BAT variability. 
Three additional cases offer multi-epoch observations, but their smooth temporal behaviour is more suggestive of an external origin.
In the remaining 15 cases, only a single-epoch optical observation is available and no hint on the origin of the optical emission can be inferred from the temporal properties.

For each analysed spectrum we extrapolated the best-fit model we obtained from the spectral analysis of XRT+BAT(+GBM) data (see the previous section) to the optical band and compared the predicted optical flux with observations.
We repeated this exercise both for synchrotron and BB+CPL fits.
The aim of our investigation was twofold. 
Firstly, we added optical data to test the consistency of the low-energy spectral index with the synchrotron value -2/3. While such a consistency would be a strong argument in favour of the synchrotron radiation, the need for a harder value would pose a severe problem for the synchrotron interpretation. 
Secondly, we tested whether optical observations might be used to distinguish between the synchrotron model and a model invoking the presence of two different spectral components, that is, a non-thermal component (CPL) plus a BB.

In order for optical observations to be a valid tool for the scopes outlined above, the radiation in the optical band has to have the same origin as the prompt radiation that is detected in the X/$\gamma$-ray band. 
We applied the test to all the 21 GRBs, but we first focus on
the three cases where the optical and the X-ray light curves are correlated. 
These are GRB\,061121, GRB\,080928, and GRB\,110205A (see Figs.~\ref{fig:061121}, \ref{fig:080928}, and \ref{fig:110205A}).
At the end of this section we discuss the results we obtained for the rest of the sample.
For these cases the optical emission cannot be firmly associated with the prompt X-ray emission, therefore they cannot be directly used to test prompt spectral models. 
However, they still allow us to infer some firm conclusions. 
Their spectral fits and comparison with optical data are shown in Appendices~\ref{app:multi} and \ref{app:single}.\\

For GRBs whose optical variability is consistent with X-ray variability, the results are shown in Figs.~\ref{fig:061121}, \ref{fig:080928}, and \ref{fig:110205A}.
In each figure the upper panel shows the light curve of the emission detected by BAT (green) and XRT (red). Optical points are shown with black symbols. 
The bottom panels collect all the time-resolved spectral energy distributions (SEDs): the synchrotron fits are shown in the left panel, and the BB+CPL fits in right panel.

In GRB\,061121, UVOT observations are available in nine epochs, corresponding to the second (and brightest) emission episode of the prompt phase. 
In GRB\,080928 the brightest emission episode begins at $\sim 170$\,s and is also observed by XRT. The UVOT began observing (in white filter) immediately after, starting at 199\,s, close to the main peak of the prompt emission. 
In GRB\,110205A, the brightest emission episodes in the prompt emission  have been observed simultaneously by BAT, XRT, and UVOT (white filter).  We note that at least two of these GRBs show a temporal lag between the optical and the X-ray light curves. Time lag is commonly measured in prompt light curves of long GRBs that are observed in different energy bands (e.g. \citealt{Cheng1995,Norris1996,Band1997}). Under the assumption that optical and X-ray emissions share the same origin, the time lag can be accounted for by the temporal evolution of the spectrum. From the inspection of the figures we draw the following conclusions that are valid for all three GRBs:
\begin{itemize}
    \item in all time-resolved spectra the optical point lies on the extrapolation of the synchrotron model;

    \item the cooling energy \ec\ ranges between 1 and 30\,keV; this implies that from the optical to the soft X-ray band, the data are fully consistent with a photon index -2/3;
    
    \item the BB+CPL model systematically overpredicts the optical flux by even one or two orders of magnitude. A low-energy break in the non-thermal component would then be mandatory to make the modelling consistent with optical observations. The result on the need of a spectral break between 1-30\,keV is then common to both models and is robust.
\end{itemize}

We note that the broadband optical-to-hard X-ray emission of GRB~110205A was previously studied by \cite{guiriec2016}, who proposed a different spectral modelling.
The joint XRT, BAT, and Suzaku/WAM spectra in their work were fitted by a three-component model, composed of two CPLs and a BB. The optical data are consistent with the extrapolation of this model. 
Three-component models are not considered in this work, as we limit our investigation to the comparison between a synchrotron model and a BB+CPL model. 

To summarise, we found that the broadband data from optical to hard X-rays are consistent with the synchrotron radiation spectrum. On the one hand, this consistency further supports the presence of a spectral break at keV energies: a simple fitting of BAT/GBM data with a Band function \citep{Band1993}, a CPL, or a smoothly broken PL model would overpredict the optical flux when it is extrapolated to the optical band. The spectral break identified in XRT data is necessary in order to match the optical data. A BB+CPL model, on the other hand, overestimates the optical flux. To be consistent with optical observations, this model would require a spectral break in the low-energy part of the non-thermal component. However, it is very likely that if such a feature is added, the BB component would no longer be needed: as demonstrated in \cite{Oganesyan2017} and \cite{ravasio2018}, the inclusion of a BB component is an alternative way of modelling the spectral break at low energies.

\subsection*{Cases with optical emission dominated by afterglow radiation}
Three additional GRBs (GRB\,070616, GRB\,081008, and GRB\,121217A) have more than one optical optical detection in the time interval of the prompt emission, allowing us to build an optical light curve and compare its temporal properties with the X-ray light curve. 
In these GRBs, the optical does not show any clear correlation with the X-ray flux. 
In contrast, the temporal behaviour is suggestive of an external origin.

Their light curves and \nufnu\ spectra fitted with synchrotron and BB+CPL models can be found in Appendix~\ref{app:multi}. 
Best-fit models are extrapolated to the optical band and compared to the optical point.
The results for these three GRBs are quite similar and can be summarised as follows:
\begin{itemize}
\item in all time-resolved spectra, consistently with the afterglow interpretation suggested by the optical light curve, the low-energy extrapolation of the synchrotron spectrum lies below the optical flux (on average one order of magnitude below);
\item in contrast, the extrapolation of the BB+CPL fit overpredicts the optical flux (requiring a break in the non-thermal component) in seven spectra and is consistent with the optical flux in the remaining four spectra.
\end{itemize}

A prominent prompt emission episode at times preceding the optical observations allows for the formation of the external shock that leads to the afterglow radiation. We note that this is indeed the case for all three GRBs: a relevant emission episode in X-ray is visible at times preceding the optical observations by at least 100\,s. An afterglow forward-shock interpretation is particularly convincing for GRB\,081008 (\ref{app:multi}). An afterglow light curve in log-scale and on a more extended temporal window can be found in \cite{Yuan2010} (see their figure 1). The optical light curve rises from 100 to 180 seconds and then decays as a PL in time, up to $10^5$\,s. The small variability around 130\,s can be interpreted as contribution from prompt radiation. This is consistent with synchrotron modelling, which predicts a 10\% contribution from prompt emission around that time.
The extrapolation of the BB+CPL fit is instead consistent with the optical flux, which is at odds with the afterglow nature of the emission. We speculate that because the BB+CPL model usually overpredicts the prompt emission in the optical, it might be by chance consistent with optical flux when the optical band is dominated by an afterglow component. However, in most of the spectra, the BB+CPL model still overpredicts the optical flux.

\subsection*{Cases with single-epoch optical observations}
The majority of GRBs in our sample (15/21 events) have a single optical exposure during the prompt emission. 
For these cases the information on the temporal evolution of the optical emission is missing and we have no indication from which we might guess the origin of the detected optical flux. 
We also built the SED from optical to hard X-ray, however, and compared them with the best-fit models, both for synchrotron and BB+CPL fits. All light curves and \nufnu\ spectra can be found in Appendix~\ref{app:single}. 

The results can be summarised as follows:
\begin{itemize}
    \item Optical fluxes are consistent with the low-energy extrapolation of the synchrotron model for 6 GRBs out of 15. 
    For the very same 6 GRBs, the BB+CPL model overpredicts the optical flux and requires a low-energy break to remove the inconsistency. These results are similar to the results obtained when the optical variability tracks the X-ray variability. This suggests that in these cases the optical emission has the same origin as the prompt X-ray emission, and the broadband spectrum is consistent with synchrotron radiation.

    \item For 6 GRBs the synchrotron model underpredicts the amount of the optical flux. In these cases the optical might be dominated by external shock radiation. Consistently with this interpretation, in all 6 cases the optical observation is preceded by a bright event in the X-ray that may be responsible for the detected optical afterglow emission. For the same 6 GRBs, the BB+CPL model either overpredicts, underpredicts, or is consistent with the optical flux.
    
    \item The remaining three cases (GRB\,090715B, GRB\,111103B, and GRB\,111123A) show inconsistency with both the synchrotron and BB+CPL models: their extrapolation overpredicts the optical flux. For the synchrotron model the discrepancy is a factor of $\lesssim$\,10 and might be explained with intrinsic absorption that is unaccounted for. For the BB+CPL model the discrepancy is much larger ($>10^2$ for GRB\,090715B, $>10$ for GRB\,111103B and GRB\,111123A).
\end{itemize}

\section{Discussion}
We have shown that synchrotron spectra provide an acceptable fit to prompt emission spectra from 0.5\,keV up to 150\,keV or higher. 
When simultaneous prompt optical observations are available and can be used to further test the spectral shape of the prompt emission, the synchrotron model continues to return acceptable fits, thus providing a good description of the radiation over four or more orders of magnitude. 
For 35 of 52 analysed spectra, spectral fits with a synchrotron function return well-constrained cooling energy \ec, \rat\ (or, equivalently, a peak energy $E_{\rm peak}\equiv(\gamma_{\rm m}/\gamma_{\rm c})^2 E_{\rm c}$), and normalisation, which has been expressed as flux $F_{\rm c}$ at the energy \ec\ (see Table~\ref{tab:table_fits}). In this section, we use these best-fit values and infer the properties of the source that are expected to give rise to synchrotron emission spectra with the observed characteristics. 

Following similar calculations performed by \citet{kumar2008} and \citet{beniamini2013}, we assume that i) the emitting region is located at distance $R_{\gamma}$ from the central engine and ii) it moves with bulk Lorentz factor $\Gamma$, iii) electrons are accelerated only once on a timescale that is negligible compared to the pulse timescale, and iv) it radiates synchrotron photons in presence of v) a constant and homogeneous magnetic field of comoving strength $B'$. 
In this simple scenario, the emission output can be fully determined by five parameters: $B'$, \gm, $\Gamma$, $R_{\gamma}$ , and $N_{\rm e}$, the latter being the number of radiating electrons.
Constraints on the physical parameters describing the emitting region are provided by the observables derived from spectral fits: \ec, \ep, and $F_{\rm c}$.
A fourth observable, the pulse decay timescale $t_{\gamma}$ , is introduced. 
To summarise, we have five unknowns and four observables. This implies that all physical quantities can be expressed as a function of one parameter, which we choose to be the bulk Lorentz factor $\Gamma$.

With respect to previous works performing a similar analysis, we have the advantage of the clear identification of the cooling frequency, which we found to lie quite close to the peak energy. A similar regime (called marginally-fast cooling) was investigated both by \citet{kumar2008} and \citet{beniamini2013} by imposing $\nu_{\rm c}\sim\nu_{\rm m}$. The interest in the marginally-fast cooling regime was motivated by its potential of solving the problem of the hard low-energy photon index that is commonly observed in GRB spectra \citep{daigne2011}.

The four observables can be expressed as a function of the five unknowns by the following equations:
\begin{equation}
E_{\rm peak} = \frac { 3\,q_{\rm e}\,h\,B^\prime\,\gamma_{\rm m}^2} {4\,\pi\,m_{\rm e}\,c} \frac{\Gamma}{1 + z}
\label{eq:Ep}
\end{equation}
\begin{equation}
E_{\rm c} = \frac { 27\,\pi \,q_{\rm e}\,h\,m_{\rm e}\,c\,(1+z)} {\sigma_{\rm T}^{2}\,B^{\prime3}\,t_{\rm \gamma}^{2}\,\Gamma (1+Y)^{2}}~,
\label{eq:Ec}
\end{equation}
\begin{equation}
F_{\rm c} = \frac {\sqrt{3}\, q_{\rm e}^3 \, B^\prime \, N_{\rm e} \, \Gamma (1+z)} {4\,\pi\,d_{\rm L }^2 \,m_{ \rm  e } \, c^2}~,
\label{eq:Fc}
\end{equation}
\begin{equation}
t_{\rm \gamma} = \frac{R_{\gamma}(1+z)}{2\,c\,\Gamma^{2}}
\label{eq:Rg}
,\end{equation}
where $z$ and $d_{\rm L}$ are redshift and luminosity distance\footnote{We adopted a standard cosmology with with $H_0= 69.7$ km s$^{-1}$Mpc$^{-1}$, $\Omega_\Lambda=0.7$, $\Omega_{\rm m}=0.3$}, $q_{\rm e}$ and $m_{\rm e}$ are electron charge and mass, and $Y$ is the Compton parameter. 
For an extensive discussion of the relation between the typical timescale and radius (Eq.~\ref{eq:Rg}), see \citealt{kumar2008}. 

We solved these equations and derived the model parameters as a function of $\Gamma$. We limited the search to $\Gamma<2000$.
Moreover, we required that the amount of SSC radiation radiation is (conservatively) less than ten times the energy output in synchrotron radiation, that is, we imposed $Y<10$, where $Y$ is the Compton parameter and is estimated as $Y \sim  \frac{4}{3} \tau \langle\gamma^2\rangle \xi_{\rm KN}$, with $\tau = N_{\rm e}\sigma_{\rm T}/ 4 \pi R_{\gamma}^2$. The factor $\xi_{KN}$ is introduced to apply a rough correction to the Compton parameter because of the  Klein-Nishina (KN) regime. We approximate $\xi_{KN}\approx \left(\frac{\gamma_{\rm KN}}{\gamma_m}\right)^{1/2}$ , when $\gamma_{\rm KN}<\gamma_m$, and $\xi_{KN}=1$ otherwise  \citep{Ando2008}, with $\gamma_{\rm KN} = m_{\rm e} c^{2} \Gamma / E_{\rm syn}(1+z)$.
The average electron Lorentz factor is computed as
\begin{equation}
\langle\gamma^{2}\rangle = \gamma_{\rm m}^{2} \frac{\gamma_{\rm c}}{\gamma_{\rm m}} \frac{ \frac{1-p}{2-p} - \frac{\gamma_{\rm c}}{\gamma_{\rm m}} }{1+\frac{\gamma_{\rm c}}{\gamma_{\rm m}} \frac{1-p}{p}}
.\end{equation}
\begin{figure}
  \includegraphics[scale=0.467]{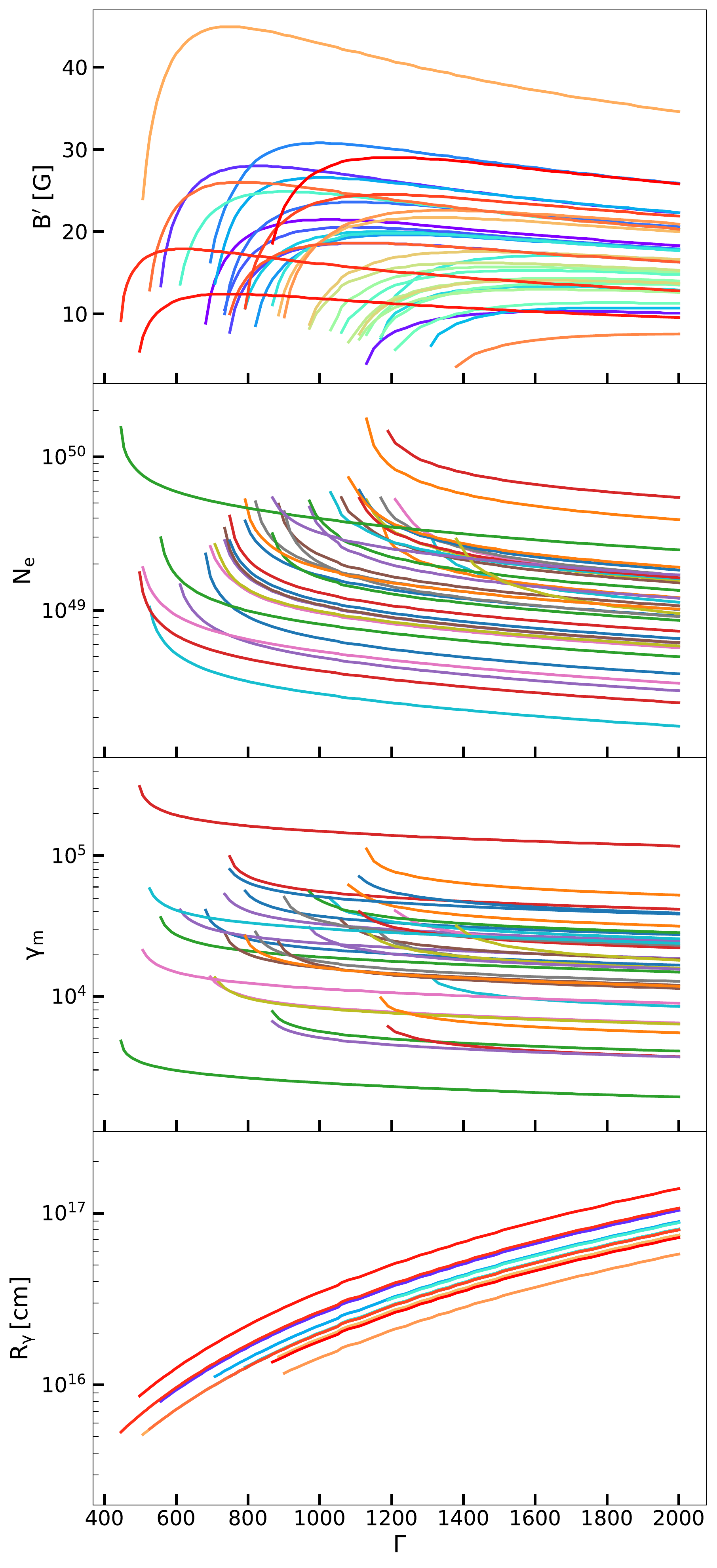}
\caption{\label{fig:constrains} Physical quantities describing the emitting region are show as a function of the bulk Lorentz factor $\Gamma$: the magnetic field strength in the comoving frame $B'$ (top panel), the number of emitting electrons $N_{\rm e}$ (second panel), the minimum Lorentz factor \gm\ of the electron (third panel), and the distance $R_{\gamma}$ of the emitting region from the central engine (bottom panel). At small $\Gamma$, solutions are limited  by the requirement $Y<10$. The pulse decay timescale $t_{\gamma}$ is fixed to 1\,s.} 
\end{figure}

Solving all equations and imposing $Y<10$ and $\tau<1,$ we infer the allowed range of values for each free model parameter as a function of $\Gamma$.
Solutions are shown in Fig.~\ref{fig:constrains} for $t_\gamma =1$\,s, for 35 time-resolved spectra for which both \ec\ and \rat\ are constrained. 
For GRBs without measured redshift, we assumed $z=2$. The lack of solutions for low values of $\Gamma$ is determined by the requirement $Y<10$. 
The allowed range of values for the physical quantities isthe following: 5\,G\,$<B'<40$\,G, $2\times10^{48}<N_{\rm e} < 10^{50}$, $2\times10^4<\gamma_{\rm m} < 10^5$, $10^{16} \, \rm cm <R_{\rm \gamma} < 10^{17}$\,cm, and $\Gamma>450$. 

These constraints are in good agreement with those derived for a regime of marginally-fast cooling by \citet{kumar2008} and \citet{beniamini2013} using similar calculations and by \citet{daigne2011} by means of numerical calculations. 
This is expected because the average ratio $\nu_{\rm m}/\nu_{\rm c}\sim10-30$ inferred from our spectral fits does not introduce any major difference as compared to the case $\nu_{\rm m}/\nu_{\rm c}\sim1$ that was investigated in those previous analyses.

As compared to the standard values that are generally invoked for the prompt emitting region, the values we have inferred point to much larger radii, a much weaker magnetic field, quite large $\Gamma$ (e.g. \citealt{Ghirlanda2018}), and relatively low values of $N_{\rm e}$. The inferred location of the emitting region is very similar to the typical deceleration radius $R_{\rm dec}$. 
To quantify a possible inconsistency between the inferred radii for the production of prompt emission and the onset of the afterglow, we estimated for each GRB the deceleration radius, assuming both a homogeneous and a wind-like density profile of the circumburst medium. To estimate the blastwave energy, we used the prompt energy and assumed a 20\% efficiency for the prompt mechanism. The only free parameter for the estimate of the deceleration radius is then the density $n$.
For each GRB, we imposed that the minimum value of $R_\gamma$ allowed by the inferred parameter space is lower than the deceleration radius and derived the upper limits on the ambient density. For a homogeneous medium we derived the upper limits of the densities in the range $1-10^{3}$\,cm$^{-3}$.
For a wind-shaped medium described by a density profile $n(R)=3\times10^{35}$\,cm$^{-1}\,A_\star\,R^{-2}$, the limits are more stringent, with values of $A_\star$ in the range $10^{-3}-1$.

This parameter space was derived assuming $t_\gamma =1$\,s. 
A shorter timescale ($t_\gamma =0.1$\,s) would imply a smaller prompt emission radius ($10^{15}$\,cm$<R_{\rm \gamma} < 2\times10^{16}$\,cm). As a consequence, the solutions in this case would point to larger bulk Lorentz factors ($\Gamma>600$) to avoid the inverse Compton scattering, larger magnetic field strengths (20\,G\,$<B'<200$\,G), allowed by the shorter timescale, a smaller number of emitting electrons ($3\times10^{47}<N_{\rm e} < 3\times10^{49}$) and of their energies ($8\times10^2<\gamma_{\rm m} < 10^5$). Longer timescales ($t_\gamma =10$\,s), conversely, imply larger radii ($2\times10^{16}$\,cm $<R_{\rm \gamma} < 10^{18}$\,cm), smaller bulk Lorentz factors ($\Gamma>330$) and magnetic fields (1\,G\,$<B'<10$\,G), a larger number of emitting electrons ($8\times10^{48}<N_{\rm e} < 9\times10^{50}$) and of their energies ($4\times10^4<\gamma_{\rm m} < 7\times10^6$).

These constraints are derived under very basic assumptions. Modifications to the allowed parameter space can be introduced by considering less standard scenarios. 
Relaxing the main underlying assumption that the observed radiation is synchrotron emission and allowing for SSC solutions implies a much more reasonable parameter space \citep{kumar2008}, but would also imply a prompt optical flux that is higher than observed. 
In the context of synchrotron scenarios, proposed solutions include a decaying magnetic field  \citep{peer2006,derishev2007,zhao2014,uhm2014,geng2018}.
However, this solution requires invoking efficient dissipation of the jet magnetic field.
Other scenarios invoke modifications to particle acceleration, as in slow particle heating \citep{asano09}, adiabatic stochastic acceleration \citep{xu2018}, or particle re-acceleration \citep{kumar2008}.

\section{Conclusions}
Previous investigations of a sample of 34 GRBs with XRT prompt observations showed that a spectral break is commonly present at energies between 1 and 20\,keV \citep{gor2018}. Below the break energy the spectrum is described by a PL with a photon index that on average is consistent with -2/3, which is reminiscent of synchrotron radiation. 
To further test and investigate the consistency of the observed spectra with synchrotron emission, in this work we have extended the analysis to even lower frequencies by including simultaneous optical observations that are available for a subsample of 21 GRBs.
We have modelled the joint XRT+BAT (and GBM data when available) time-resolved spectra of all 21 GRBs and extrapolated the best-fit model to the optical band, where it was compared with the optical flux.
We performed this analysis using two competing models: synchrotron and BB+CPL. 
The synchrotron model was built adopting the synchrotron kernel and assuming that the electron distribution is the result of a PL injection spectrum modified by synchrotron and inverse Compton cooling.
The results can be summarised as follows: 

\begin{itemize}
\item The synchrotron model provides a good description of all prompt emission spectra (see the $\chi^{2}$ distribution in Fig.~\ref{fig:chisq}).
In almost all cases, the best fit is obtained for synchrotron radiation in the marginally-fast cooling regime, that is, the cooling frequency is located at keV energies (see Fig.~\ref{fig:syn_results}, right panel). The distribution of the cooling energy peaks at \ec\,$\sim 3$\,keV. 
The ratio between the minimum Lorentz factor of the injected electron population and the cooling Lorentz factor is on average \rat\,$\sim$\,4 (Fig.~\ref{fig:syn_results}, left panel).

\item Similarly good fits are obtained with a BB+CPL model (Fig.~\ref{fig:chisq}). The BB component peaks at $\sim$\,5\,keV, and its presence is required in all but two spectra.\\
\end{itemize}

These results show that as long as only X-ray spectral data are considered, both models are acceptable. Optical observations are fundamental to distinguishing among the two. From the analysis of optical observations we conclude that the synchrotron model is strongly favoured. To summarise our findings, we divided the sample into three subsamples, according tho the temporal properties of the optical flux.

The results for the comparison of optical observations and synchrotron fits can be summarised as follows:
\begin{itemize}

\item 
In three GRBs the optical and X-ray light curves show temporally correlated variability. These cases offer a test bench for prompt spectral models because the correlated variability strongly suggests that optical and X/$\gamma$-ray radiation share the same origin.
In all these GRBs (and all their time-resolved spectra), the optical flux lies on the extrapolation of the synchrotron fits. The spectrum is thus consistent with having photon index -2/3 from optical to soft X-rays.

\item In three GRBs the multi-epoch optical observations show a temporal behaviour that is suggestive of an external origin. Consistently, the optical flux is well in excess of the flux predicted by extrapolating the synchrotron best-fit model of the prompt emission.

\item For the remaining 15 GRBs, only a single-epoch optical observation is available. We find that in 6 cases the optical flux lies on the extrapolation of the synchrotron model, pointing to an internal origin and to a synchrotron interpretation of the broadband emission. In an additional 6 cases the optical flux lies well above the synchrotron extrapolation, supporting an interpretation of the optical in terms of external shock radiation. This latter interpretation is consistent with the presence of a major emission episode in X-rays at times preceding the optical observation \citep{Nappo2014}. In the remaining 3 GRBs the synchrotron model overpredicts the optical flux by a factor of $\le 10$. 
The probable solution of this discrepancy is an unaccounted-for intrinsic absorption in the host galaxy.

\end{itemize}

Although GRBs with single-epoch observations and GRBs with a smooth optical light curve do not provide a strong test for the synchrotron model (but only a consistency check), they allow us to draw robust conclusions for the BB+CPL fits.

The results on the BB+CPL fits can be summarised as follows:
\begin{itemize}

\item The BB+CPL model overpredicts the amount of optical flux for most of the analysed prompt emission spectra. In particular, in spite of the large errors in the extrapolation of the model (much larger than in the case of the synchrotron model), the BB+CPL fits i) overpredict (at more than 3$\sigma$) the optical flux in 34 of 52 spectra (65\%), ii) give marginally consistent (between 1 and 3$\sigma$) values in 7 cases, and iii) are consistent within 1$\sigma$ in 10 cases. 

\item In particular, for the three GRBs where the optical and X-ray light curves show temporally correlated variability, we note that the BB+CPL model overpredicts the optical flux by a large factor (on average between 10 and 10$^2$);

\end{itemize}

Most BB+CPL fits, even though they provide a good fit to soft/hard X-ray data, are then inconsistent with (i.e. overpredict) the optical flux. To remove this inconsistency, a spectral break between the optical and the X-ray band needs to be introduced.
However, as demonstrated by \cite{Oganesyan2017,gor2018}, the inclusion of a low-energy break in the empirical fitting functions for this sample of GRBs would remove the need for a BB component. In principle, the CPL+BB model can be reconciled with optical observations if the optical flux is strongly attenuated by unaccounted-for intrinsic absorption in their host galaxy. However, this would require visual extinctions  
of $A_V$>1, which is observed for less than 10\% of the GRB population \citep{Covino2013}. \\

Motivated by the encouraging results we obtained from the synchrotron fits, we have assumed that the dominant radiative process is synchrotron radiation, and we investigated which constraints can be set on the physical parameters of the source.
A regime of marginally-fast cooling, such as is supported by our results, has previously been considered by a few authors \citep{kumar2008,daigne2011,beniamini2013} to relax the inconsistency between the predictions from fast-cooling synchrotron radiation and the somewhat harder spectra observed in GRBs. As pointed out in these studies, the condition $\nu_{\rm c}\sim \nu_{\rm m}$ implies non-standard values of the physical parameters.
According to our analysis, the inferred values for the comoving magnetic field, the number of emitting electrons, the electron Lorentz factor, the emitting radius, and the bulk Lorentz factor are 5\,G$<B'<45$\,G, $2\times10^{48}<N_{\rm e} < 10^{50}$, $2\times10^4<\gamma_{\rm m} < 10^5$, $R_{\rm \gamma} >5\times10^{15}$\,cm, and $\Gamma>450$, respectively.

The results presented in this work are very encouraging for the identification of the dominant radiative mechanism in the prompt emission of GRBs. On the other hand, this step forward immediately faces great difficulty when it is used to infer the properties of the emitting region. A synchrotron emission in marginally-fast cooling does not easily fit in a standard scenario for prompt emission. 
Several proposals have been put forward in the past to save a marginally-fast cooling regime and derive more reasonable values for the physical parameters describing the region where the emission is produced. No obvious solution has been identified.
The investigation of advantages and drawbacks of these models will be the focus of a paper in preparation.

\begin{acknowledgements}
LN acknowledges funding from the European Union's Horizon 2020 Research and Innovation programme under the Marie Sk\l odowska-Curie grant agreement n.\,664931. G.G. acknowledges Prin-INAF 2017 1.05.01.88.06. We acknowledge ASI grant 1.05.04.3.19. This research has been supported by ASI grant INAF I/004/11/1. This research has made use of data obtained through the High Energy Astrophysics Science Archive Research Center Online Service, provided by the NASA/Goddard Space Flight Center, and specifically, this work made use of public \fe-GBM data. This work made use of data supplied by the UK Swift Science Data Centre at the University of Leicester.

\end{acknowledgements}

\bibliographystyle{aa} 
\bibliography{references} 

\begin{appendix}
\onecolumn 
\section{Optical observations}\label{app:table_optical}

\begin{longtable}{ccccc}
\caption{\label{tab:table_optical} Optical observations: time interval (observer frame, since the trigger time), flux ($F_{\rm opt}$), and optical filter. The notation 'G' or 'G+HG' (fourth column) clarifies whether the correction for extinction has been applied considering only the Galactic contribution or also the contributions from the host galaxy. The reference for the observations is provided in the last column.}\\
\hline\hline \\
Time interval & F$_{\rm opt}$ & Filter  &Ext. & Ref.\\ [0.01 cm]
   s          & {\rm mJy}     &       &   &     \\ [0.1 cm]
\hline 
\endfirsthead 

\caption{continued.}\\
\hline\hline\\
Time interval & F$_{\rm opt}$ & Filter  &Ext. & Ref.\\ [0.01 cm]
   s          & {\rm mJy}     &       &   &     \\ [0.1 cm]
\hline \\
\endhead

\hline
\endfoot    

\\
\multicolumn{5}{c}{${ \bf GRB~060510B }$ } \\   [0.1 cm]
\hline\\
$180.00 - 210.00$  &$0.03952\pm 0.01591$  & R &G & \citet{omelandri2008} \\
\\
\multicolumn{5}{c}{${ \bf GRB~060814 }$ } \\   [0.1 cm]
\hline\\
$123.00 - 213.00$  & $<0.4297$ & r& G & \citet{oklotz2006} \\
\\
\multicolumn{5}{c}{${ \bf GRB~061121 }$ } \\   [0.1 cm]
\hline \\
$63.99 - 68.83$  & $0.2797\pm 0.0361$&White& G+HG & \citet{Page2007} \\ 
$68.83 - 71.89$  & $0.5547 \pm 0.0707$ \\ 
$71.89 - 74.89$  & $0.5748 \pm 0.0728$ \\ 
$74.90 - 76.96$  & $1.041\pm 0.1539$\\
$76.96 - 79.00$  & $1.0370\pm 0.1547$ \\ 
$79.00 - 81.54$  & $0.7476\pm 0.09889$ \\ 
$81.54 - 84.68$  & $0.5386\pm 0.0675$\\
$84.68 - 89.68$  & $0.2613\pm 0.03461$ \\
$89.69 - 95.79$  & $0.1961\pm 0.02681$\\ 
\\
\multicolumn{5}{c}{${ \bf GRB~070616 }$ } \\   [0.1 cm]
\hline\\
$250.30 - 385.30$  & $0.3112\pm 0.105$ & V & G & \citet{starling2008} \\ 
$385.40 - 448.00$  & $1.4597\pm 0.154$\\ 
$448.20 - 509.60$  & $1.7352\pm 0.155$\\ 
$509.70 - 572.10$  & $1.5534\pm 0.155$\\ 
$572.30 - 631.50$  & $2.3411\pm 0.160$ \\ 
\\
\multicolumn{5}{c}{${ \bf GRB~080928 }$ } \\   [0.1 cm]
\hline\\
$199.00 - 219.00$  & $0.1488\pm 0.0346$& $\rm R_C$ & G+HG & \citet{rossi2011} \\ 
$219.00 - 239.00$  & $0.2016 \pm 0.0392$ \\ 
$239.00 - 259.00$  & $0.2562 \pm 0.0403$\\ 
$259.00 - 278.70$  & $0.1171 \pm 0.0339$\\
\\
\multicolumn{5}{c}{${ \bf GRB~081008 }$ } \\   [0.1 cm]
\hline\\
$104.81 - 109.81$  & $0.8832$ & r & G+HG & \citet{Yuan2010} \\ 
$113.34 - 118.34$  & $1.2598$\\ 
$120.96 - 125.96$  & $1.7969$\\
$127.41 - 132.41$  & $2.2769$\\ 
\\
\multicolumn{5}{c}{${ \bf GRB~090715B }$ } \\   [0.1 cm]
\hline\\
$53.10 - 200.10$  & $0.0266\pm 0.0037$ & White & G & \citet{ovetere2009} \\ 
\\
\multicolumn{5}{c}{${ \bf GRB~100906A}$ } \\   [0.1 cm]
\hline\\
$116.30 - 136.30$  & $22.3\pm 2.057$ & R & G+HG & \citet{Gorbovskoy2012} \\ 
\\
\multicolumn{5}{c}{${ \bf GRB~110102A }$ } \\   [0.1 cm]
\hline\\
$156.00 - 303.00$  & $ 0.1496\pm 0.0083$& White& G & \citet{ooates2011} \\ 
\\
\multicolumn{5}{c}{${ \bf GRB~110119A }$ } \\   [0.1 cm]
\hline\\
$67.00 - 214.00$  & $0.6121   \pm 0.0169$ & White & G & \citet{opritchard2011} \\  
\\
\multicolumn{5}{c}{${ \bf GRB~110205A }$ } \\   [0.1 cm]
\hline\\
 $163.50 - 168.50$  & $0.0686\pm 0.0218$ & White & G & \citet{ocucchiara2011} \\ 
 $168.50 - 173.50$  & $0.0581 \pm 0.0208$ \\ 
 $173.50 - 178.50$  & $0.0620\pm 0.0221$ \\ 
 $181.00 - 191.00$  & $0.0745\pm 0.0216$\\ 
 $198.50 - 203.50$  & $0.0692\pm 0.0227$ \\ 
 $203.50 - 208.50$  & $0.0794\pm 0.0230$ \\ 
 $208.50 - 213.50$  & $0.2089\pm 0.0329$ \\ 
 $213.50 - 218.50$  & $0.1585\pm 0.0279$ \\ 
 $218.50 - 223.50$  & $0.1117\pm 0.0260$ \\ 
 $223.50 - 228.50$  & $0.0824\pm 0.0223$\\ 
 $228.50 - 233.50$  & $0.0847\pm 0.0237$ \\ 
 $234.00 - 248.00$  & $0.0745\pm 0.0223$ \\ 
 $251.00 - 281.00$  & $0.0643\pm 0.0217$ \\ 
\\
\multicolumn{5}{c}{${ \bf GRB~111103B }$ } \\   [0.1 cm]
\hline\\
$68.00 - 215.00$  & $<0.0204$ & White & G & \citet{ooates2011b} \\ 
\\
\multicolumn{5}{c}{${ \bf GRB~111123A }$ } \\   [0.1 cm]
\hline\\
$110.00 - 257.00$  & $<0.007178$ & White & G & \citet{oholland2011} \\ 
\\
\\
\multicolumn{5}{c}{${ \bf GRB~121123A }$ } \\   [0.1 cm]
\hline\\
$131.00 - 278.00$  & $0.0368\pm 0.0051$ & White & G & \citet{oholland2012} \\ 
\\
\multicolumn{5}{c}{${ \bf GRB~121217A }$ } \\   [0.1 cm]
\hline\\
$598.50 - 639.50$  & $0.1823\pm 0.0151$ & H & G & \citet{elliott2014} \\ 
  &  $0.1003\pm 0.0130$ & K  \\ 
  &  $0.1985\pm 0.0238$ & J \\ 
$723.50 - 764.50$  & $0.1926\pm 0.0107$ & H & &  \\
  &  $0.1195\pm 0.0110$ & K \\
  &  $0.2217\pm 0.0184$ & J \\
\\
\multicolumn{5}{c}{${ \bf GRB~130514A }$ } \\   [0.1 cm]
\hline\\
$106.20 - 166.20$  & $ 0.2357\pm 0.0437 $ & r & G & \citet{oschmidl2013} \\ 
\\
\multicolumn{5}{c}{${ \bf GRB~130907A }$ } \\   [0.1 cm]
\hline
\\
$266.00 - 306.00$  & $1.069\pm 0.0394$ & u & G+HG & \citet{overes2015} \\ 
\\
\multicolumn{5}{c}{${ \bf GRB~140108A }$ } \\   [0.1 cm]
\hline
\\
$78.00 - 225.00$  & $0.1141\pm 0.0211$ & White & G & \citet{obreeveld2014} \\
\\
\multicolumn{5}{c}{${ \bf GRB~140206A}$ } \\   [0.1 cm]
\hline
\\
$52.00 - 199.00$  & $1.178\pm 0.0217$ & White & G & \citet{ooates2014} \\ 
\\
\multicolumn{5}{c}{${ \bf GRB~140512A }$ } \\   [0.1 cm]
\hline
\\
$126.00 - 146.00$  & $24.49\pm 0.203$ & R& G+HG & \citet{ohuang2016} \\ 
\\
\multicolumn{5}{c}{${ \bf GRB~151021A }$ } \\   [0.1 cm]
\hline
\\
$111.90 - 141.90$  & $5.664\pm 1.049$ & I & G & \citet{otrotter2015} \\ 
\\
\end{longtable}
\vskip 2.5truecm


\section{Synchrotron and CPL+BB fits}\label{app:table_fits}

\begin{longtable}{c|cccc|cccc}
\caption{\label{tab:table_fits} Best-fit parameters of the synchrotron and BB+CPL models for the time-resolved spectra. The table lists the time interval (since BAT trigger time), the best-fit parameters (cooling energy \ec, \rat, flux $\rm F_{\rm c}$ at the peak of the \fnu\ spectrum, total chi-square $\chi^{2}$ and degrees of freedom (d.o.f.). For BB+CPL fits the table lists the CPL photon index $\alpha$, the CPL peak energy \ep, the BB temperature $kT$, $\chi^{2}$ , and the d.o.f.. }\\
\hline
&\multicolumn{4}{|c}{Synchrotron model}& \multicolumn{4}{|c}{CPL+BB model}\\
\hline
 Time interval & $E_{\rm c}$ & $\gamma_{\rm m}/\gamma_{\rm c}$ & $F_{\rm c}$ &   $\rm \chi^{2}$ (d.o.f.) & $\alpha$ & $E_{\rm p}$ & $kT$  & $\rm \chi^{2}$ (d.o.f.) \\
     ${\rm s}$ & ${\rm keV}$ &                                 &  ${\rm mJy}$   & &  & ${\rm keV}$  & ${\rm keV}$  & \\
\hline
\multicolumn{9}{c}{}
\endfirsthead 

\caption{continued.}\\
\hline
&\multicolumn{4}{|c}{Synchrotron model} &\multicolumn{4}{|c}{CPL+BB model}\\
\hline
 Time interval & $E_{\rm c}$ & $\gamma_{\rm m}/\gamma_{\rm c}$ & $F_{\rm c}$ &   $\rm \chi^{2}$ (d.o.f.) & $\alpha$ & $E_{\rm p}$ & $kT$  & $\rm \chi^{2}$ (d.o.f.) \\
     ${\rm s}$ & ${\rm keV}$ &                                 &  ${\rm mJy}$   & &  & ${\rm keV}$  & ${\rm keV}$  \\
     \hline \\
\endhead

\hline
\endfoot    

\multicolumn{9}{c}{${ \bf GRB~060510B }$ } \\   [0.1 cm]
\hline\\
 $180.00 - 210.00$  & $2.5_{-0.4}^{+0.5}$ & $4.8_{-1.7}^{+1.8}$ & $0.33 \pm 0.01$ & 188 (196) & $-1.18_{-0.08}^{+0.09}$ & $56.4_{-5.7}^{+6.4}$ & $1.01_{-0.15}^{+0.15}$  &189 (194) \\ 
\\
\multicolumn{9}{c}{${ \bf GRB~060814 }$ } \\   [0.1 cm]
\hline\\
 $123.00 - 213.00$  & $<0.2$ & $29.3_{-11.9}^{+4.2}$ & $>1.60$  &292 (298) & $-1.53_{-0.07}^{+0.07}$ & $115.9_{-16.0}^{+17.6}$ & $0.31_{-0.06}^{+0.07}$  &252 (296) \\
\\
\multicolumn{9}{c}{${ \bf GRB~061121 }$ } \\   [0.1 cm]
\hline\\
$63.99 - 68.83$  & $14.3_{-4.2}^{+3.5}$ & $>20.8$ & $2.42 \pm 0.10$ & 113 (87) & $-1.11_{-0.07}^{+0.09}$ & $256.1_{-19.2}^{+25.7}$ & $2.55_{-0.74}^{+1.21}$  &92 (85) \\
$68.83 - 71.89$  & $10.7_{-3.6}^{+2.8}$ & $6.3_{-1.5}^{+4.0}$ & $4.02 \pm 0.21$ & 76 (67) & $-0.97_{-0.08}^{+0.06}$ & $111.8_{-8.5}^{+6.8}$ & $2.65_{-0.34}^{+0.58}$  &80 (65) \\  
$71.89 - 74.89$  & $24.1_{-3.6}^{+3.4}$ & $>13.5$ & $5.24 \pm 0.13$ & 83 (74) & $-0.99_{-0.10}^{+0.07}$ & $338.5_{-34.1}^{+22.9}$ & $6.47_{-2.18}^{+1.58}$  &69 (72) \\ 
$74.90 - 76.96$  & $32.7_{-5.0}^{+5.0}$ & $>6.6$ & $4.06 \pm 0.11$ & 69 (68) & $-1.09_{-0.07}^{+0.12}$ & $383.9_{-31.2}^{+49.1}$ & $2.24_{-1.29}^{+7.81}$  &51 (66) \\ 
$76.96 - 79.00$  & $3.6_{-1.7}^{+2.0}$ & $>9.1$ & $2.90 \pm 0.33$ & 69 (61) & $-1.31_{-0.14}^{+0.12}$ & $99.3_{-19.7}^{+16.8}$ & $1.69_{-0.75}^{+0.71}$  &65 (59) \\ 
$79.00 - 81.54$  & $<1.6$ & $12.1_{-5.3}^{+38.8}$ & $>3.33$ & 61 (68) & $-1.56_{-0.05}^{+0.18}$ & $179.5_{-20.2}^{+71.9}$ & $5.05_{-0.00}^{+0.00}$  &60 (66) \\  
$81.54 - 84.68$  & $<1.1$ & $13.9_{-6.0}^{+34.5}$ & $>2.22$ & 80 (79) & $-1.32_{-0.21}^{+0.27}$ & $78.0_{-23.8}^{+31.4}$ & $0.50_{-0.17}^{+0.26}$  &78 (77) \\ 
$84.68 - 89.68$  & $1.1_{-0.7}^{+0.9}$ & $8.3_{-3.6}^{+9.3}$ & $1.16 \pm 0.12$ & 74 (80) & $-1.40_{-0.10}^{+0.27}$ & $76.0_{-13.1}^{+34.4}$ & $1.05_{-0.40}^{+1.58}$  &73 (79) \\ 
$89.69 - 95.79$  & $<1.6$ & $31.8_{-25.3}^{+21.5}$ & $>4.13$ & 102 (86) & $-1.42_{-0.12}^{+0.08}$ & $71.1_{-14.7}^{+9.9}$ &  &99 (86) \\ 
\\
\multicolumn{9}{c}{${ \bf GRB~070616 }$ } \\   [0.1 cm]
\hline\\
$250.30 - 385.30$  & $3.5_{-0.3}^{+0.7}$ & $9.6_{-2.0}^{+3.3}$ & $0.56 \pm 0.01$ & 304 (301) & $-1.21_{-0.03}^{+0.07}$ & $111.2_{-4.1}^{+9.7}$ & $1.84_{-0.33}^{+0.57}$  &296 (299) \\  
$385.40 - 448.00$  & $2.7_{-0.4}^{+0.4}$ & $6.8_{-1.1}^{+2.2}$ & $0.52 \pm 0.01$ & 223 (241) & $-1.24_{-0.04}^{+0.09}$ & $74.4_{-3.9}^{+8.9}$ & $1.38_{-0.37}^{+0.36}$  &212 (239) \\ 
$448.20 - 509.60$  & $1.8_{-0.3}^{+0.4}$ & $4.4_{-0.9}^{+1.0}$ & $0.60 \pm 0.02$ & 150 (186) & $-1.37_{-0.09}^{+0.04}$ & $46.3_{-6.6}^{+2.6}$ & $1.24_{-0.22}^{+0.36}$  &140 (184) \\  
$509.70 - 572.10$  & $0.9_{-0.3}^{+0.3}$ & $3.7_{-0.8}^{+1.9}$ & $0.69 \pm 0.05$ & 234 (234) & $-1.48_{-0.13}^{+0.08}$ & $24.8_{-6.1}^{+3.6}$ & $0.84_{-0.24}^{+0.38}$  &232 (232) \\  
$572.30 - 631.50$  & $<0.3$ & $8.6_{-3.5}^{+1.8}$ & $>1.74$ & 260 (226) & $-1.80_{-0.11}^{+0.15}$ & $13.2_{-7.4}^{+10.2}$ & $2.83_{-2.83}^{+7.38}$  &246 (224) \\  
\\
\multicolumn{9}{c}{${ \bf GRB~080928 }$ } \\   [0.1 cm]
\hline\\
$199.00 - 219.00$  & $2.5_{-0.5}^{+0.6}$ & $3.9_{-0.8}^{+1.1}$ & $1.15 \pm 0.02$ & 137 (137) & $-1.39_{-0.04}^{+0.17}$ & $78.6_{-5.4}^{+21.6}$ & $1.56_{-0.13}^{+0.36}$  &113 (135) \\  
$219.00 - 239.00$  & $1.3_{-0.3}^{+0.4}$ & $3.0_{-0.8}^{+2.0}$ & $0.83 \pm 0.03$  & 127 (143) & $-1.38_{-0.12}^{+0.15}$ & $30.3_{-5.7}^{+7.4}$ & $0.80_{-0.16}^{+0.21}$  &129 (141) \\ 
$239.00 - 259.00$  & $<0.4$ & $10.9_{-5.8}^{+7.7}$ & $>1.41$ & 114 (108) & $-1.69_{-0.10}^{+0.13}$ & $24.6_{-8.1}^{+10.4}$ & $0.70_{-0.22}^{+0.34}$  &109 (106) \\ 
$259.00 - 278.70$  & $<0.2$ & $10.1_{-4.8}^{+4.7}$ & $>1.04$ & 120 (122) & $-1.65_{-0.09}^{+0.10}$ & $10.0_{-2.6}^{+3.0}$ &   &118 (122) \\ 

\\
\multicolumn{9}{c}{${ \bf GRB~081008 }$ } \\   [0.1 cm]
\hline\\
$104.81 - 109.81$  & $11.6_{-6.8}^{+43.4}$ & $0.9_{-0.5}^{+1.6}$ & $0.58 \pm 0.03$ & 81 (72) & $-0.99_{-0.06}^{+0.15}$ & $35.7_{-2.2}^{+5.4}$ & $1.64_{-0.78}^{+1.47}$  &80 (70) \\ 
$113.34 - 118.34$  & $2.7_{-0.6}^{+2.0}$ & $7.6_{-3.3}^{+10.5}$ & $0.79 \pm 0.05$ & 57 (73) & $-1.18_{-0.07}^{+0.23}$ & $78.0_{-6.8}^{+21.6}$ & $1.18_{-0.88}^{+5.26}$  &54 (71) \\ 
$120.96 - 125.96$  & $6.8_{-4.6}^{+25.5}$ & $0.8_{-0.6}^{+2.1}$ & $0.74 \pm 0.06$ & 68 (74) & $-1.10_{-0.24}^{+0.22}$ & $27.0_{-7.2}^{+6.5}$ & $1.03_{-0.33}^{+1.28}$  &66 (72) \\ 
$127.41 - 132.41$  & $2.8_{-1.6}^{+3.4}$ & $1.1_{-0.7}^{+1.4}$ & $0.74 \pm 0.04$ & 74 (92) & $-1.71_{-0.10}^{+0.22}$ & $18.9_{-6.7}^{+14.6}$ & $1.12_{-0.22}^{+0.35}$  &70 (90) \\  

\\
\multicolumn{9}{c}{${ \bf GRB~090715B }$ } \\   [0.1 cm]
\hline\\
$53.10 - 200.10$  & $<0.2$ & $18.7_{-2.1}^{+3.2}$ & $>1.11$ & 363 (329) & $-1.61_{-0.04}^{+0.04}$ & $66.7_{-7.3}^{+7.1}$ & $0.58_{-0.09}^{+0.10}$  &321 (327) \\

\\
\multicolumn{9}{c}{${ \bf GRB~100906A}$ } \\   [0.1 cm]
\hline\\
$116.30 - 136.30$  & $2.9_{-0.8}^{+3.5}$ & $1.0_{-0.7}^{+0.6}$ & $5.53 \pm 0.32$ & 211 (179)   & $-1.55_{-0.17}^{+0.23}$ & $9.7_{-3.7}^{+4.9}$ & $1.06_{-0.16}^{+0.17}$  &193 (178) \\  

\\
\multicolumn{9}{c}{${ \bf GRB~110102A }$ } \\   [0.1 cm]
\hline\\
$156.00 - 303.00$  & $1.9_{-0.1}^{+0.1}$ & $>57.1$ & $0.87 \pm 0.01$ & 425 (305) & $-1.27_{-0.04}^{+0.02}$ & $185.9_{-10.5}^{+4.9}$ & $4.92_{-2.56}^{+2.05}$  &343 (303) \\ 

\\
\multicolumn{9}{c}{${ \bf GRB~110119A }$ } \\   [0.1 cm]
\hline\\
$67.00 - 214.00$  & $1.8_{-0.2}^{+0.2}$ & $6.6_{-0.9}^{+1.3}$ & $0.29 \pm 0.00$ & 421 (475) & $-1.24_{-0.05}^{+0.05}$ & $67.9_{-4.3}^{+4.6}$ & $0.89_{-0.07}^{+0.08}$  &430 (473) \\ 

\\
\multicolumn{9}{c}{${ \bf GRB~110205A }$ } \\   [0.1 cm]
\hline\\
$163.50 - 168.50$  & $4.9_{-1.2}^{+2.8}$ & $3.3_{-1.1}^{+1.8}$ & $0.89 \pm 0.04$ & 82 (78) & $-1.11_{-0.11}^{+0.09}$ & $66.7_{-7.9}^{+6.9}$ & $1.84_{-0.49}^{+0.55}$  &67 (76) \\ 
 $168.50 - 173.50$  & $10.4_{-3.6}^{+37.0}$ & $2.4_{-1.5}^{+1.6}$ & $0.70 \pm 0.03$ & 76 (78) & $-0.98_{-0.09}^{+0.32}$ & $66.7_{-5.7}^{+20.7}$ & $2.46_{-1.13}^{+0.90}$  &74 (76) \\ 
 $173.50 - 178.50$  & $6.2_{-1.6}^{+2.6}$ & $2.4_{-1.0}^{+1.3}$ & $0.90 \pm 0.04$ & 102 (77) & $-1.04_{-0.16}^{+0.16}$ & $53.8_{-8.7}^{+8.7}$ & $1.64_{-0.43}^{+0.96}$  &92 (75) \\  
 $181.00 - 191.00$  & $5.6_{-1.1}^{+1.6}$ & $2.6_{-0.8}^{+0.9}$ & $0.86 \pm 0.03$ & 120 (106) & $-1.00_{-0.09}^{+0.13}$ & $53.5_{-4.8}^{+6.8}$ & $1.29_{-0.23}^{+0.32}$  &104 (104) \\  
 $198.50 - 203.50$  & $4.7_{-1.1}^{+1.3}$ & $4.8_{-1.5}^{+2.8}$ & $0.95 \pm 0.04$ & 83 (77) & $-1.10_{-0.07}^{+0.05}$ & $82.9_{-6.9}^{+4.3}$ & $2.31_{-0.75}^{+0.43}$  &77 (75) \\ 
 $203.50 - 208.50$  & $5.6_{-1.2}^{+1.3}$ & $>5.8$ & $1.03 \pm 0.04$ & 83 (79) & $-1.02_{-0.07}^{+0.12}$ & $104.9_{-7.1}^{+12.3}$ & $2.11_{-0.90}^{+0.57}$  &72 (77) \\ 
 $208.50 - 213.50$  & $9.9_{-3.3}^{+2.4}$ & $>7.6$ & $1.00 \pm 0.04$ & 74 (77)  & $-0.99_{-0.17}^{+0.04}$ & $149.5_{-24.3}^{+6.5}$ & $3.29_{-0.87}^{+0.45}$  &73 (75) \\ 
 $213.50 - 218.50$  & $7.7_{-1.1}^{+5.0}$ & $5.2_{-2.2}^{+4.4}$ & $0.91 \pm 0.04$ & 88 (75) & $-1.05_{-0.09}^{+0.10}$ & $103.1_{-9.5}^{+10.5}$ & $3.34_{-1.37}^{+1.50}$  &93 (73) \\  
 $218.50 - 223.50$  & $6.1_{-1.7}^{+1.5}$ & $4.9_{-1.5}^{+2.8}$ & $1.02 \pm 0.05$ & 87 (76) & $-1.04_{-0.09}^{+0.12}$ & $78.3_{-7.6}^{+9.6}$ & $1.61_{-0.42}^{+0.88}$  &80 (74) \\ 
 $223.50 - 228.50$  & $4.5_{-1.2}^{+1.4}$ & $5.4_{-1.9}^{+4.2}$ & $0.81 \pm 0.04$ & 89 (75) & $-1.01_{-0.13}^{+0.13}$ & $74.7_{-9.4}^{+10.0}$ & $1.20_{-0.21}^{+0.34}$  &78 (73) \\  
 $228.50 - 233.50$  & $6.0_{-1.4}^{+0.8}$ & $3.5_{-1.2}^{+0.8}$ & $0.87 \pm 0.03$ & 85 (77) & $-1.04_{-0.09}^{+0.12}$ & $61.5_{-5.9}^{+7.9}$ & $1.72_{-0.46}^{+0.66}$  &78 (75) \\  
 $234.00 - 248.00$  & $3.5_{-0.8}^{+1.0}$ & $3.5_{-1.0}^{+1.3}$ & $0.78 \pm 0.03$ & 128 (101) & $-1.20_{-0.05}^{+0.09}$ & $60.8_{-3.8}^{+6.9}$ & $1.35_{-0.25}^{+0.17}$  &119 (99) \\  
 $251.00 - 281.00$  & $2.0_{-0.5}^{+0.6}$ & $3.7_{-1.0}^{+1.2}$ & $0.84 \pm 0.03$ & 143 (117) & $-1.34_{-0.09}^{+0.07}$ & $64.3_{-8.7}^{+7.1}$ & $1.18_{-0.19}^{+0.33}$  &133 (115) \\ [0.2 cm]

\\
\multicolumn{9}{c}{${ \bf GRB~111103B }$ } \\   [0.1 cm]
\hline\\
$68.00 - 215.00$  & $0.4_{-0.1}^{+0.1}$ & $9.2_{-1.5}^{+2.1}$ & $0.55 \pm 0.02$ & 194 (167) & $-1.42_{-0.08}^{+0.07}$ & $46.6_{-6.1}^{+5.7}$ & $0.45_{-0.06}^{+0.07}$  &174 (165) \\ 

\\
\multicolumn{9}{c}{${ \bf GRB~111123A }$ } \\   [0.1 cm]
\hline\\
$110.00 - 257.00$  & $2.3_{-0.2}^{+0.2}$ & $6.4_{-0.9}^{+1.2}$ & $0.41 \pm 0.01$ & 283 (310) & $-1.13_{-0.05}^{+0.04}$ & $59.2_{-3.2}^{+2.6}$ & $0.85_{-0.08}^{+0.13}$  &291 (308) \\ 

\\
\multicolumn{9}{c}{${ \bf GRB~121123A }$ } \\   [0.1 cm]
\hline\\
$131.00 - 278.00$  & $28.8_{-6.0}^{+51.7}$ & $1.0_{-0.4}^{+0.7}$ & $0.34 \pm 0.01$  & 231 (242) & $-0.82_{-0.02}^{+0.06}$ & $76.3_{-1.6}^{+4.1}$ & $2.21_{-1.09}^{+2.41}$  &219 (240) \\ 

\\
\multicolumn{9}{c}{${ \bf GRB~121217A }$ } \\   [0.1 cm]
\hline\\
$598.50 - 639.50$  & $1.9_{-0.5}^{+0.6}$ & $8.5_{-3.9}^{+10.0}$ & $0.17 \pm 0.02$ & 143 (134) & $-1.38_{-0.13}^{+0.12}$ & $45.3_{-9.2}^{+8.5}$ & $1.26_{-0.37}^{+1.48}$  &144 (132) \\ 
$723.50 - 764.50$  & $3.6_{-0.6}^{+0.6}$ & $8.8_{-2.4}^{+5.9}$ & $0.50 \pm 0.01$ & 204 (180) & $-1.26_{-0.09}^{+0.06}$ & $141.0_{-17.1}^{+10.6}$ & $2.04_{-0.47}^{+1.09}$  &214 (178) \\ 

\\
\multicolumn{9}{c}{${ \bf GRB~130514A }$ } \\   [0.1 cm]
\hline
\\
$106.20 - 166.20$  & $1.5_{-0.3}^{+0.3}$ & $4.9_{-0.7}^{+1.0}$ & $1.05 \pm 0.03$ & 188 (179) & $-1.37_{-0.09}^{+0.03}$ & $47.1_{-6.6}^{+2.5}$ & $3.19_{-1.93}^{+0.76}$  &175 (177) \\ 
\\
\multicolumn{9}{c}{${ \bf GRB~130907A }$ } \\   [0.1 cm]
\hline
\\
$266.00 - 306.00$  & $5.5_{-1.4}^{+1.6}$ & $0.4_{-0.1}^{+0.1}$ & $3.64 \pm 0.26$  & 211 (199) & $-1.80_{-0.08}^{+0.04}$ & $56.1_{-20.8}^{+10.2}$ & $1.46_{-0.28}^{+0.43}$  &186 (197) \\
\\
\multicolumn{9}{c}{${ \bf GRB~140108A }$ } \\   [0.1 cm]
\hline
\\
$78.00 - 225.00$  & $10.3_{-0.8}^{+0.9}$ & $40.6_{-0.0}^{+0.0}$ & $0.16 \pm 0.00$ & 428 (436) & $-1.11_{-0.04}^{+0.05}$ & $589.4_{-28.4}^{+34.3}$ & $0.67_{-0.10}^{+0.11}$  &346 (434) \\
\\
\multicolumn{9}{c}{${ \bf GRB~140206A}$ } \\   [0.1 cm]
\hline
\\
$52.00 - 199.00$  & $9.1_{-2.2}^{+0.9}$ & $>23.9$ & $0.55 \pm 0.01$  & 87 (101) & $-1.07_{-0.03}^{+0.07}$ & $137.6_{-4.2}^{+10.6}$ & $2.25_{-0.61}^{+1.09}$  &72 (99) \\ 
\\
\multicolumn{9}{c}{${ \bf GRB~140512A }$ } \\   [0.1 cm]
\hline
\\
$126.00 - 146.00$  & $10.3_{-1.4}^{+0.9}$ & $16.1_{-5.1}^{+11.8}$ & $0.80 \pm 0.02$ & 268 (233) & $-1.11_{-0.04}^{+0.03}$ & $323.8_{-13.6}^{+11.3}$ & $3.93_{-0.84}^{+0.55}$  &225 (231) \\ 
\\
\multicolumn{9}{c}{${ \bf GRB~151021A }$ } \\   [0.1 cm]
\hline
\\
$111.90 - 141.90$  & $1.0_{-0.4}^{+1.0}$ & $1.9_{-1.0}^{+1.4}$ & $1.59 \pm 0.07$ & 150 (164) & $-1.60_{-0.15}^{+0.13}$ & $13.3_{-5.1}^{+4.4}$ & $0.79_{-0.19}^{+0.19}$  &145 (162) \\ 
\\

\end{longtable}

\clearpage
\section{Light curves and spectra}\label{app:plots}
In this section, we report the light curves and time-resolved spectra for the GRBs in the sample with prompt optical emission that do not show temporal correlation with the X-ray emission. 
In particular, we collect in Appendix \ref{app:multi} the three cases whose optical temporal behaviour resembles afterglow radiation.
In Appendix \ref{app:single} we collect the 15 cases for which only one optical observation is available during the prompt phase.

\subsection{Multi-epoch optical data with smooth temporal behaviour}\label{app:multi}
\begin{figure*}[hb!]
\caption{We here report the results of the time-resolved spectral analysis of three GRBs with multi-epoch optical observations and no evident correlation between X-ray and optical variability. For each GRB the upper panel shows the XRT (red), BAT (green), GBM (blue, if available), and optical (black) light curves. The shaded grey region corresponds to the time window where the joint XRT and BAT (and GBM) spectral analysis is performed. The time-bins for the spectral analysis are defined by the optical exposure times. The synchrotron and BB+CPL fits are shown in the left and right bottom panels, respectively. For clarity purposes, spectra are shifted along the vertical axis by multiplicative factors, as specified. The XRT flux has been de-absorbed. The best-fit contour regions are shown in orange: light orange is used for the best-fit model derived when the calibration constant was fixed to BAT data, and dark orange when it was fixed to XRT data. The optical fluxes for each time-bin (in black) are added to compare with the low-energy extrapolation of the synchrotron and BB+CPL models.}
\begin{center}
   {\centering
  \includegraphics[width=0.67\textwidth]{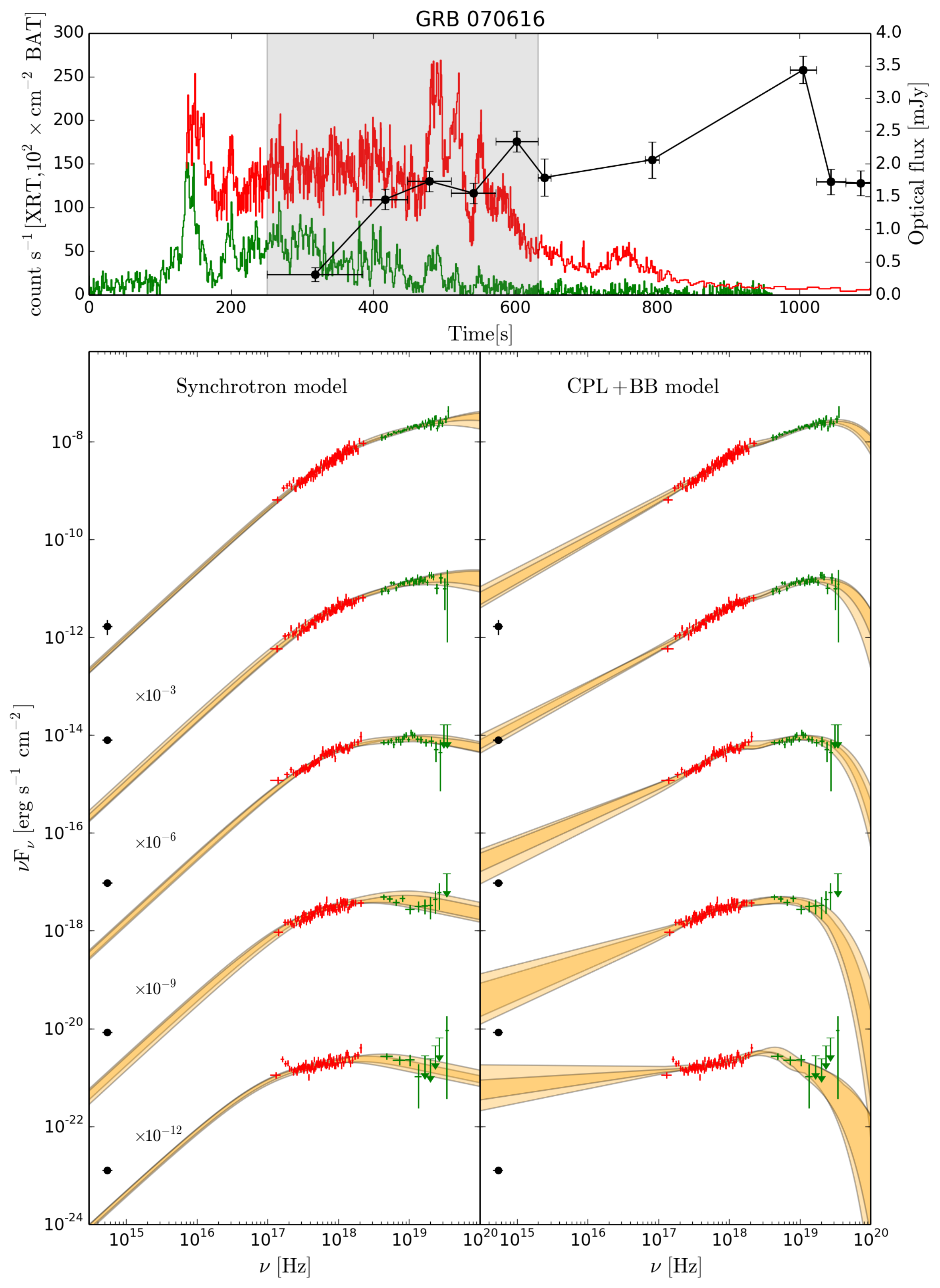}
   }
   \label{fig:070616}
\end{center}
\end{figure*}
\begin{figure*}[ht!]
\begin{center}
   {\centering
  \includegraphics[width=0.8\textwidth]{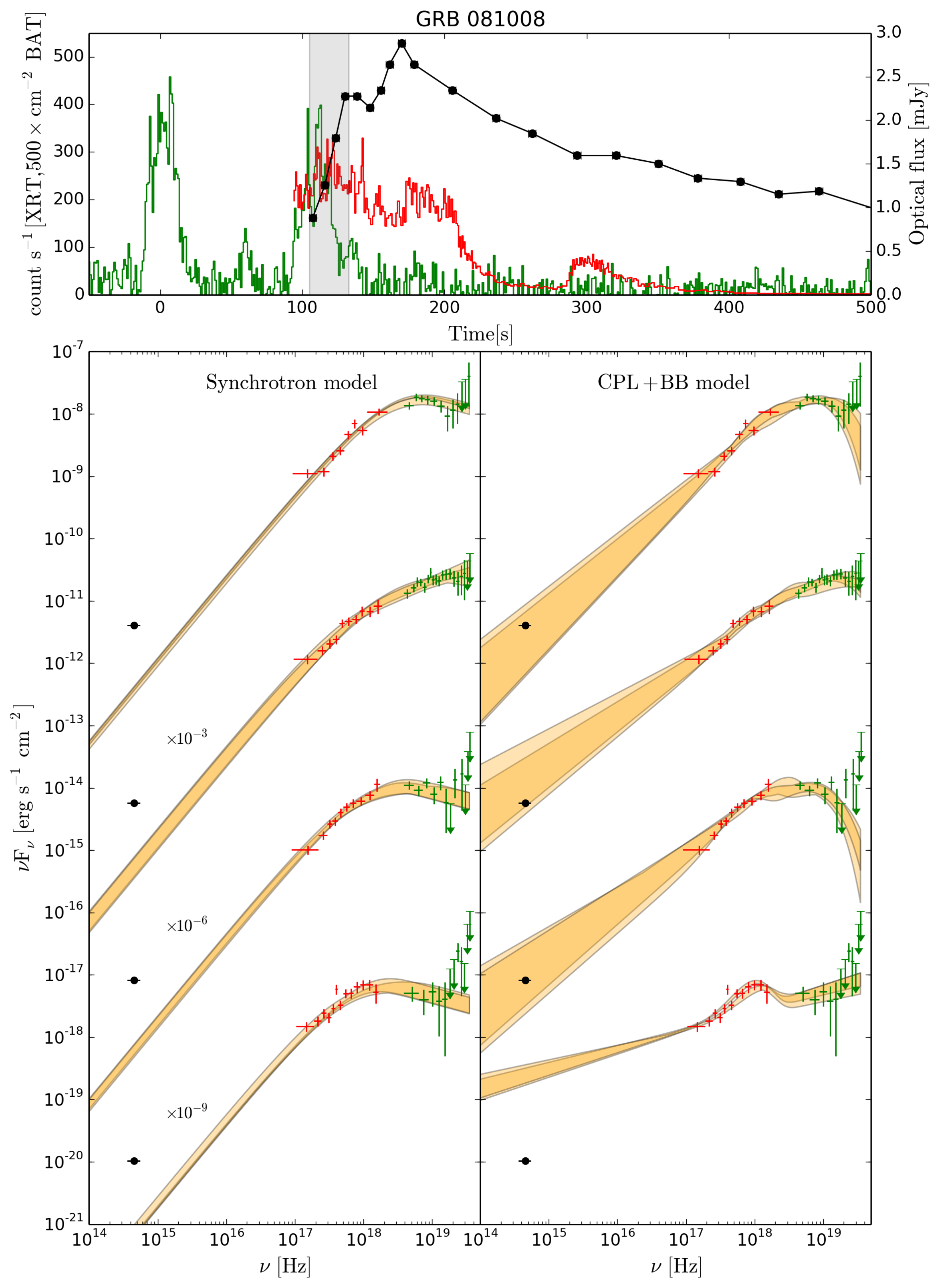}
   }
   \label{fig:081008}
\end{center}
\end{figure*}

\begin{figure*}[ht!]
\begin{center}
   {\centering
  \includegraphics[width=0.8\textwidth]{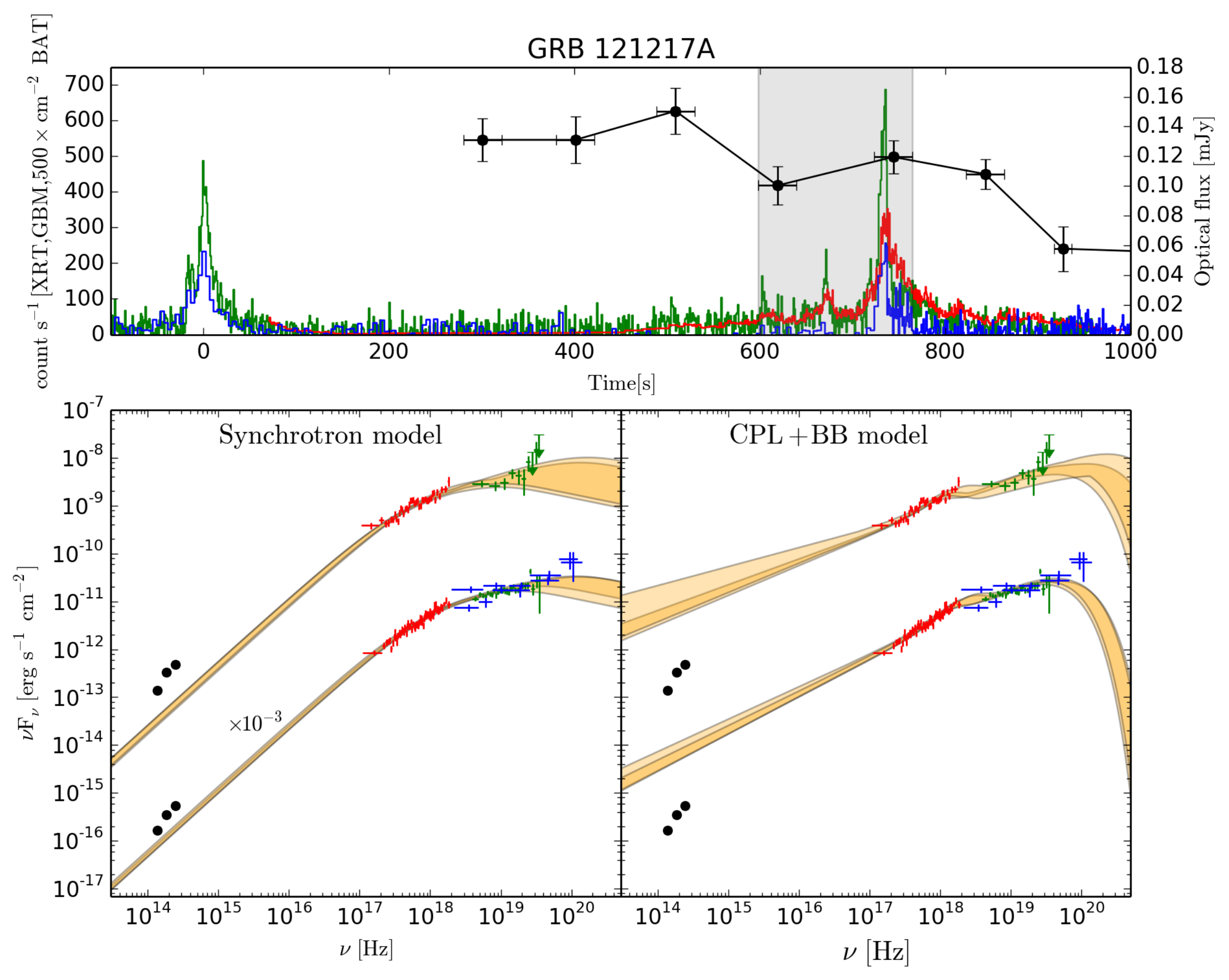}
   }
   \label{fig:121217A}
\end{center}
\end{figure*}
\clearpage
\subsection{Single-epoch optical data}\label{app:single}
\begin{figure*}[hb!]
\caption{Results of the spectral analysis of 15 GRBs with single-epoch optical observations. For each GRB the upper panel shows the XRT (red), BAT (green), and GBM (blue, if available) light curves. The black symbol shows the optical observation. The shaded grey region corresponds to the time-window of the optical exposure, where the joint XRT+BAT(+GBM) spectral analysis is performed. The synchrotron and BB+CPL fits to XRT+BAT(+GBM) spectra are shown in the middle and bottom panels, respectively. The XRT flux has been de-absorbed. The best-fit contour regions are shown in orange: light orange is used for model derived when the calibration constant in fixed on BAT data, and dark orange when it is fixed on XRT data. The optical fluxes for each time-bin (in black) are added to compare with the low-energy extrapolation of the synchrotron and BB+CPL models.}
\begin{center}
   {
  \includegraphics[width=0.48\textwidth]{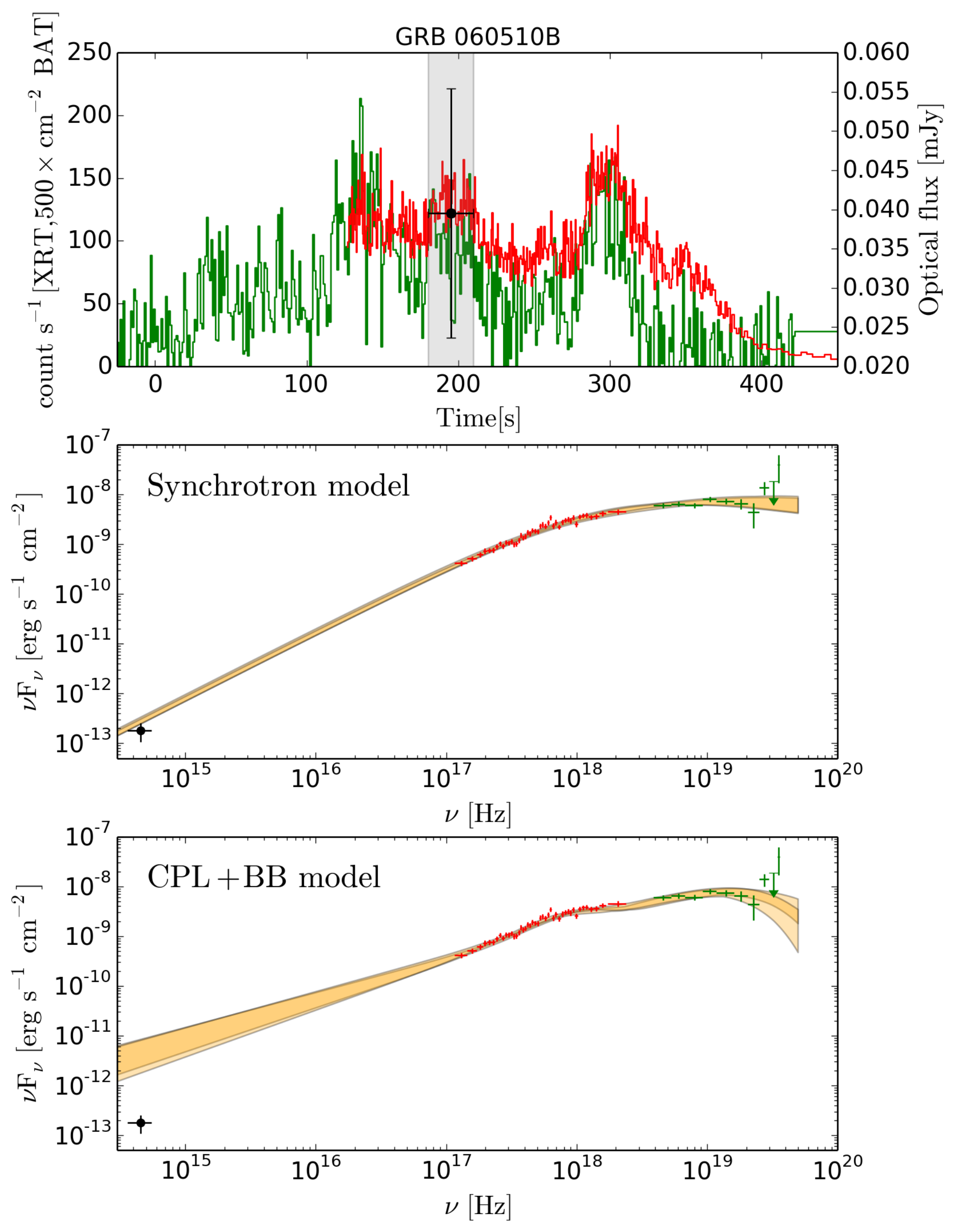}
  \hskip 0.5truecm
  \includegraphics[width=0.48\textwidth]{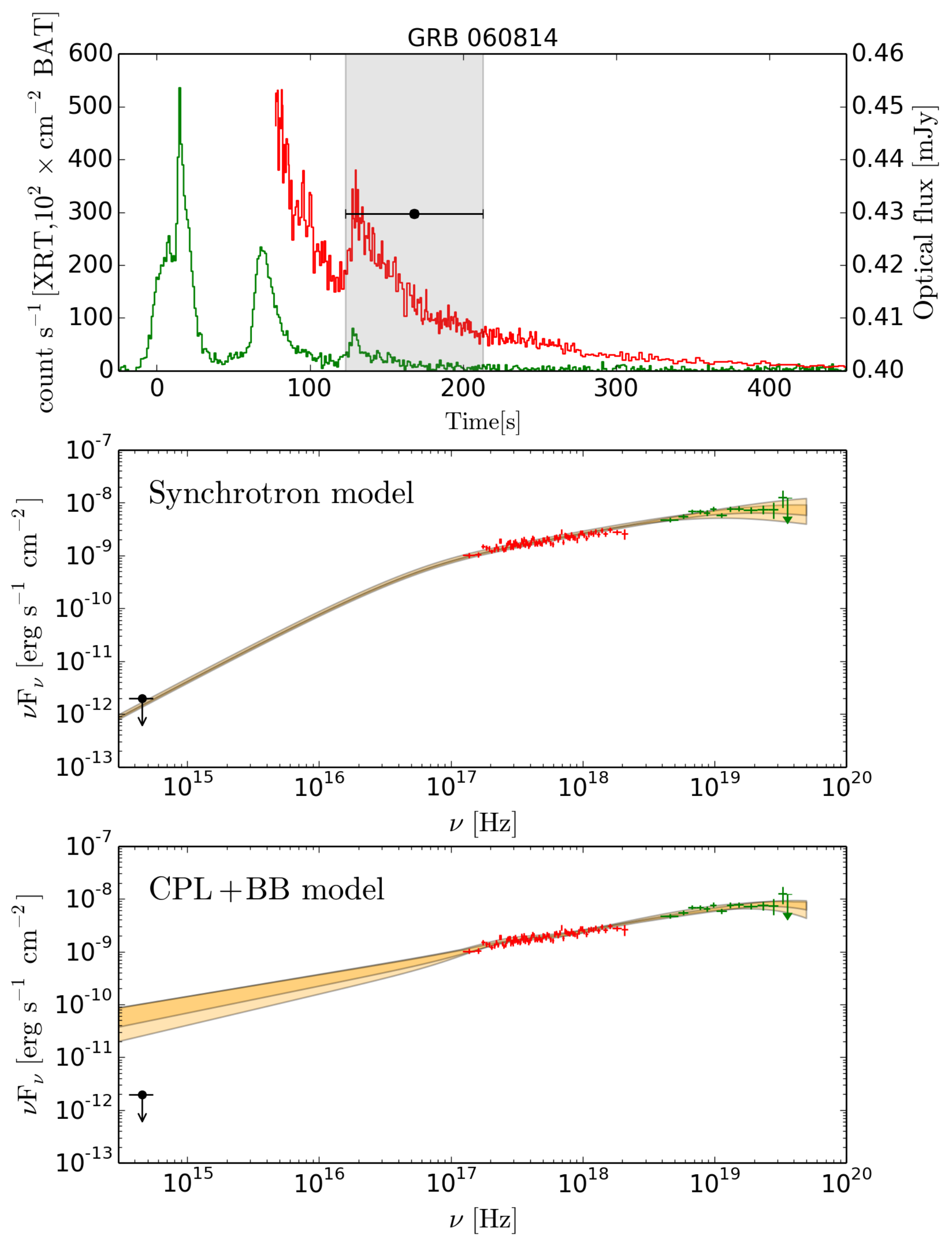}
   }
   \label{fig:singleexp}
\end{center}
\end{figure*}
\begin{figure*}[ht!]
\begin{center}
   {
   \includegraphics[width=0.48\textwidth]{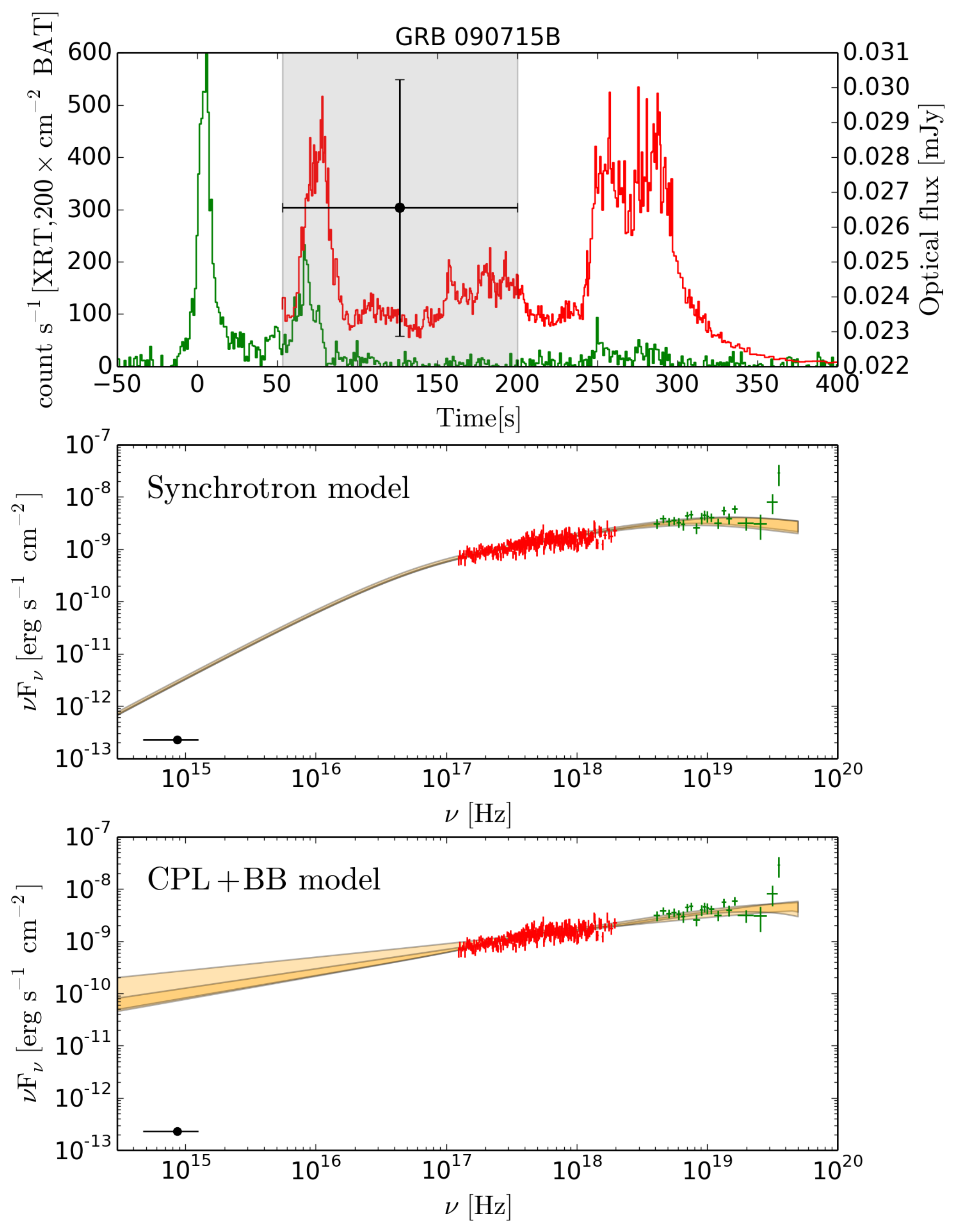}
  \hskip 0.5truecm
  \includegraphics[width=0.48\textwidth]{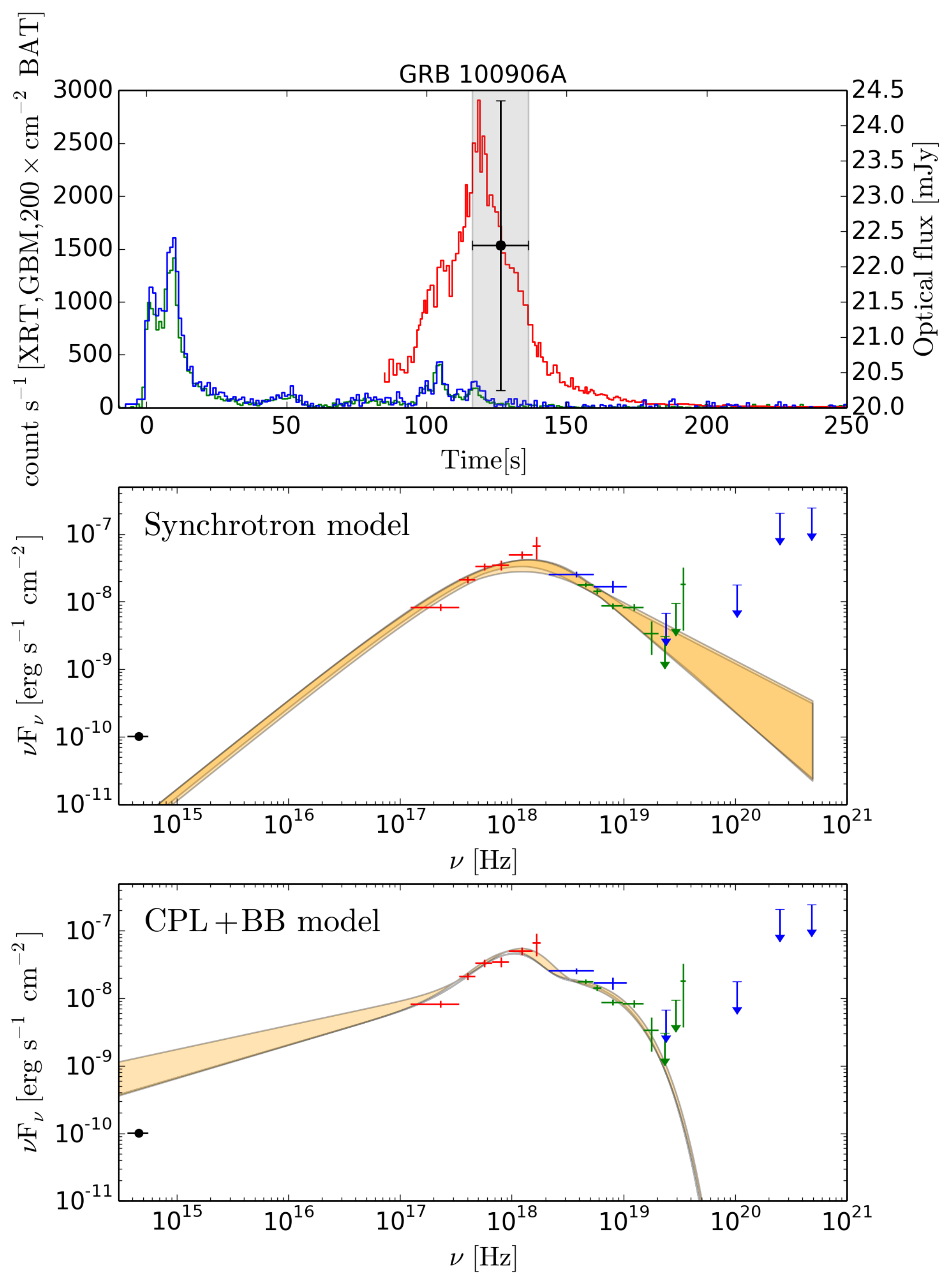}
  \vskip 0.8truecm
   \includegraphics[width=0.48\textwidth]{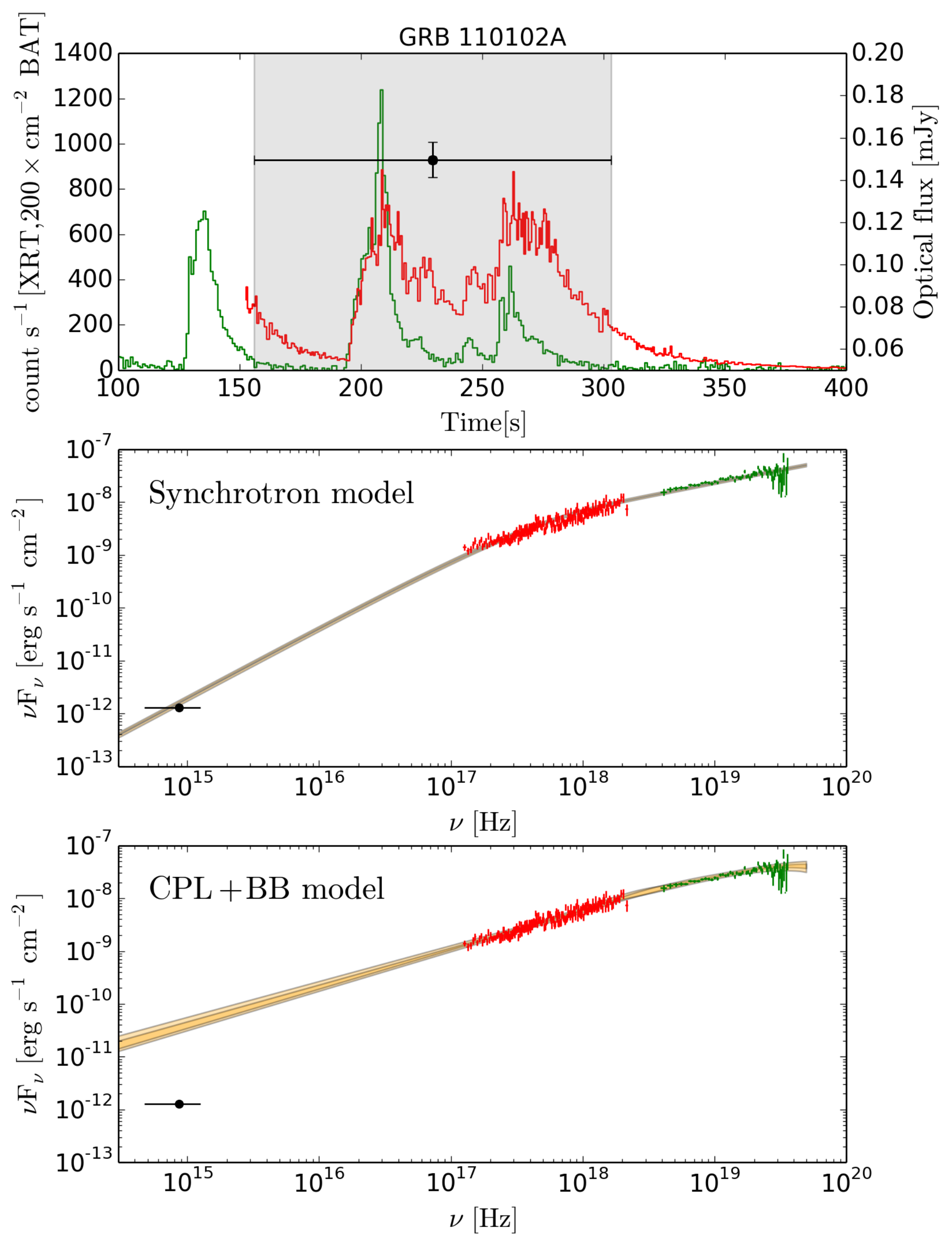}
   \hskip 0.5truecm
   \includegraphics[width=0.48\textwidth]{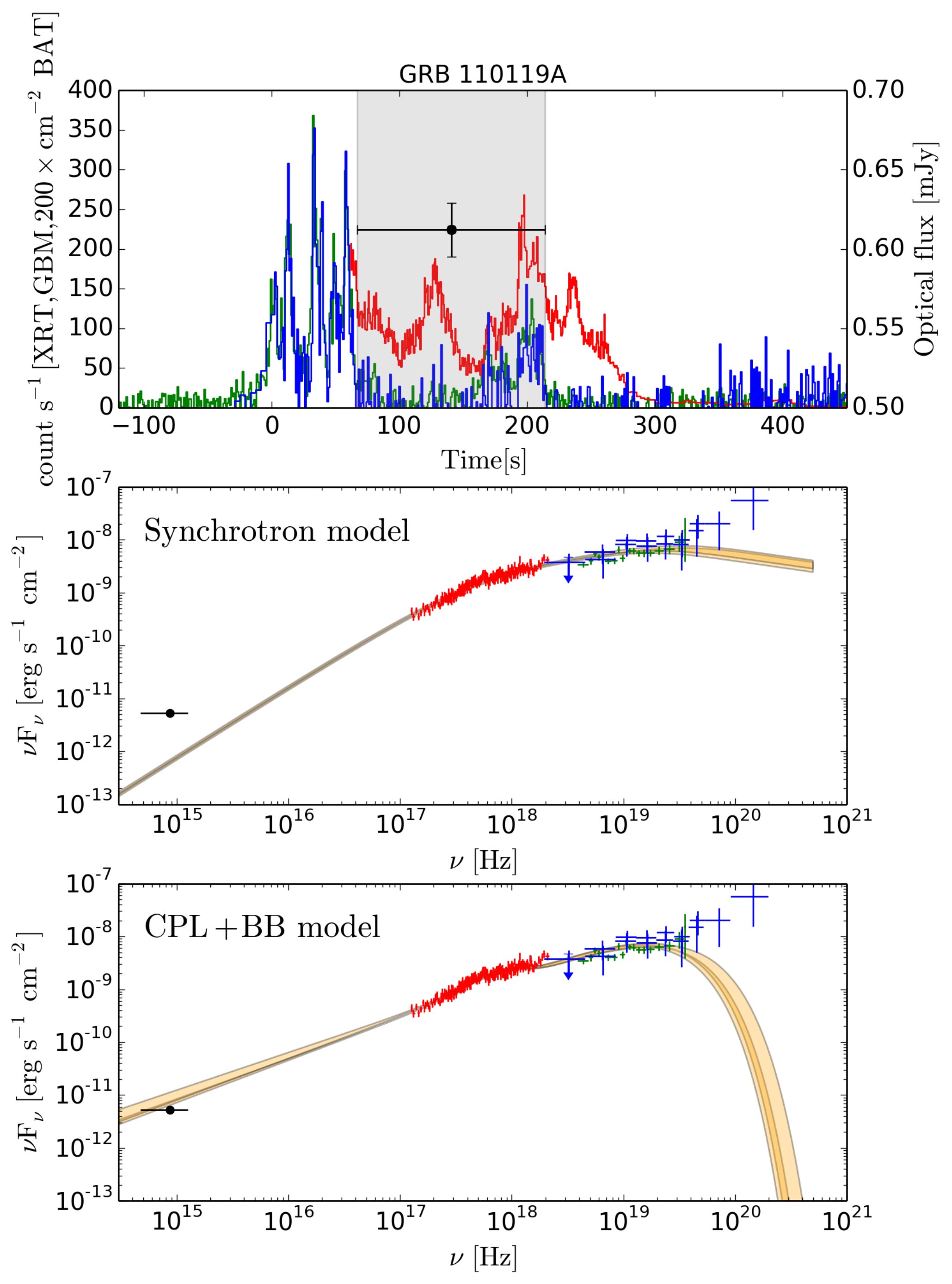}
   }
   \label{fig:singleexp1}
\end{center}
\end{figure*}

\begin{figure*}[ht!]
\begin{center}
   {
   \includegraphics[width=0.48\textwidth]{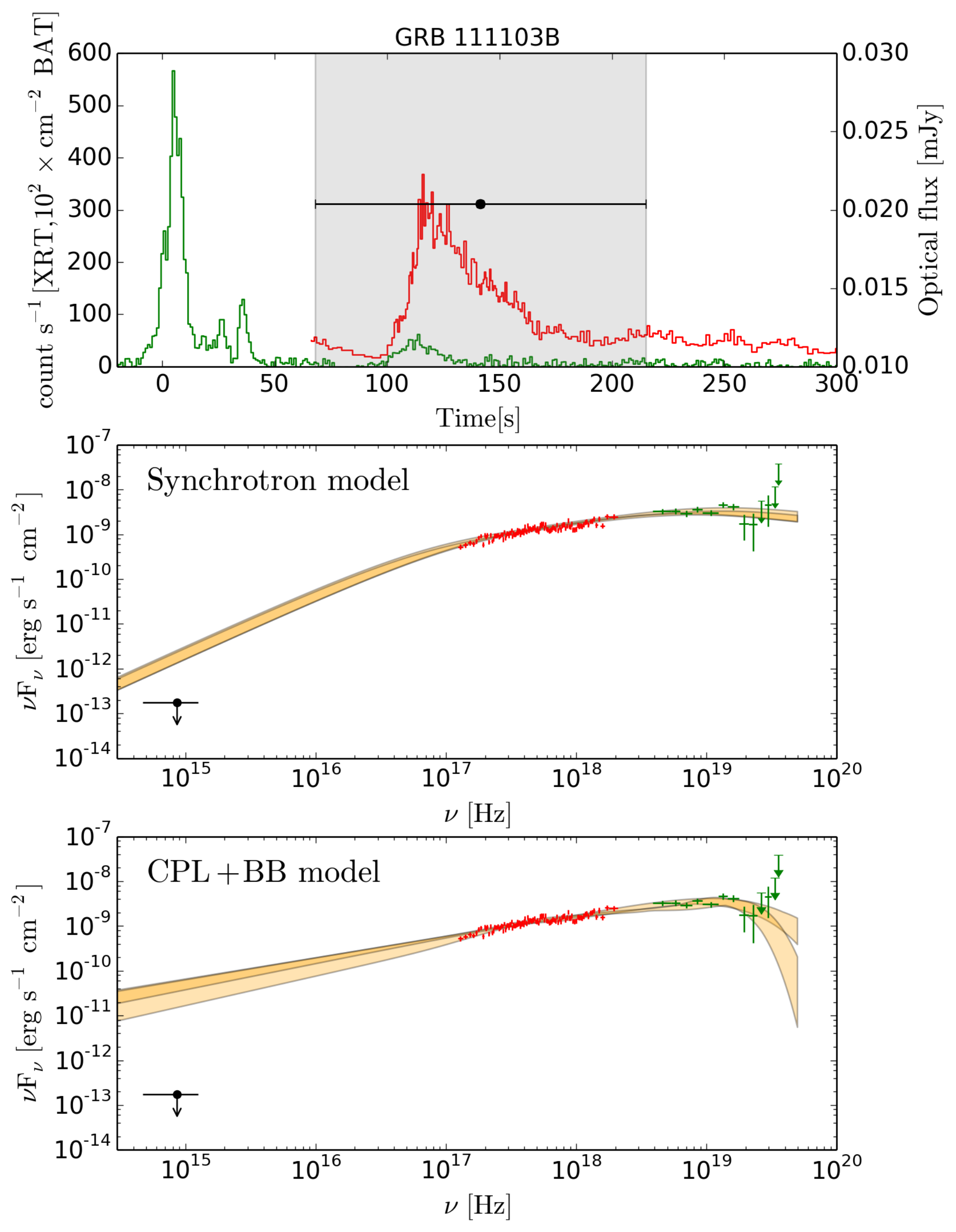}
   \hskip 0.5truecm
   \includegraphics[width=0.48\textwidth]{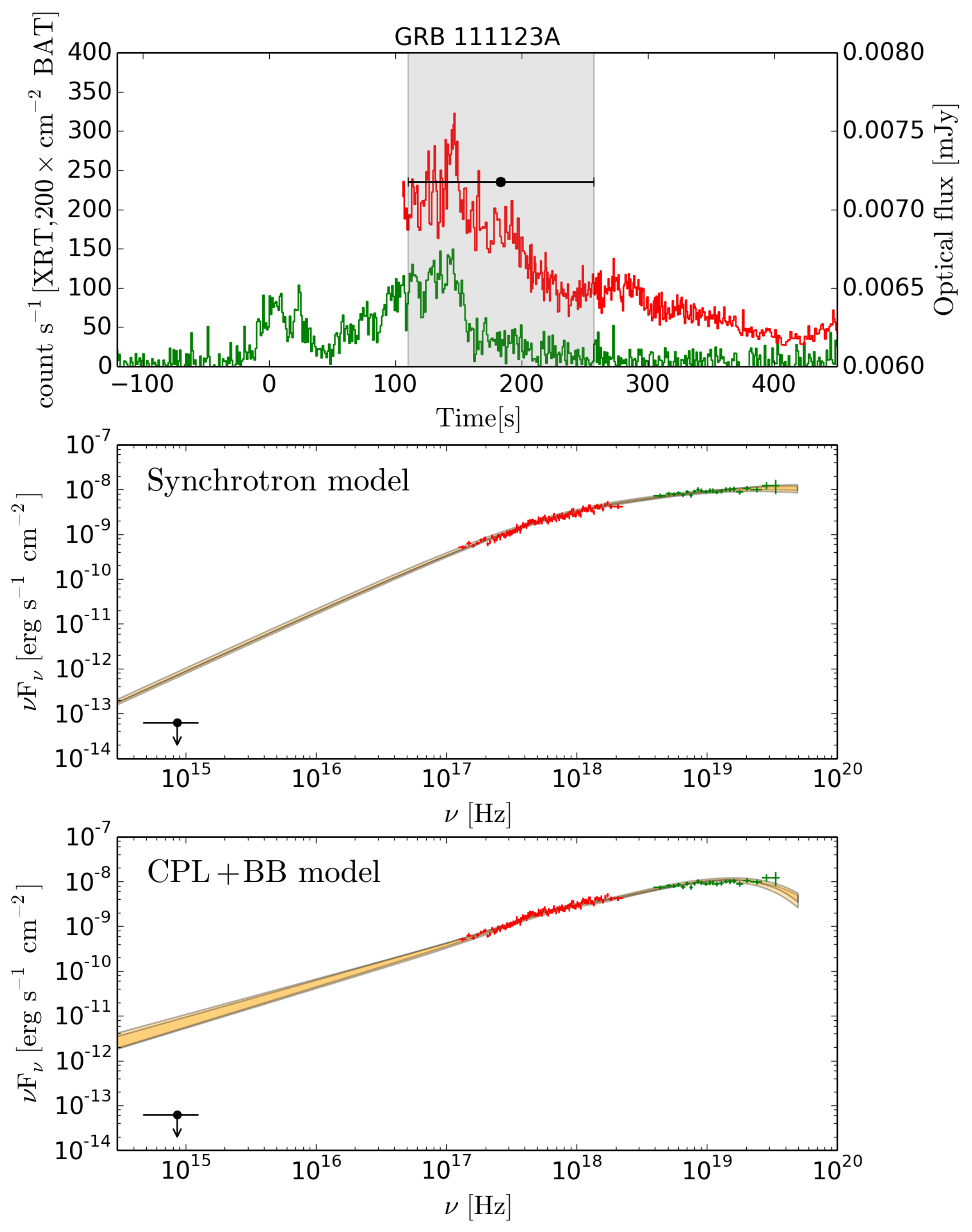}
   \vskip 0.8truecm
    \includegraphics[width=0.48\textwidth]{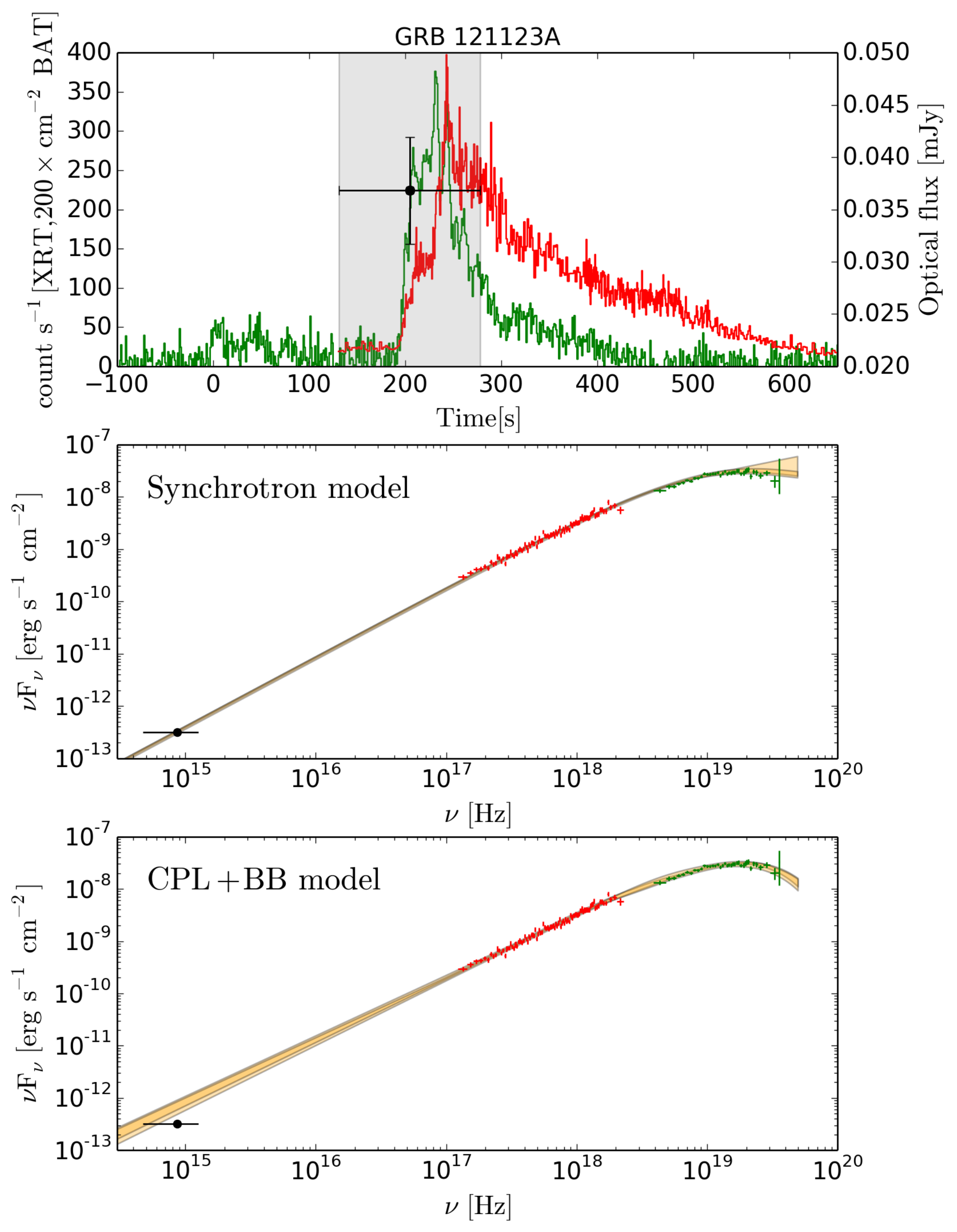}
   \hskip 0.5truecm
    \includegraphics[width=0.48\textwidth]{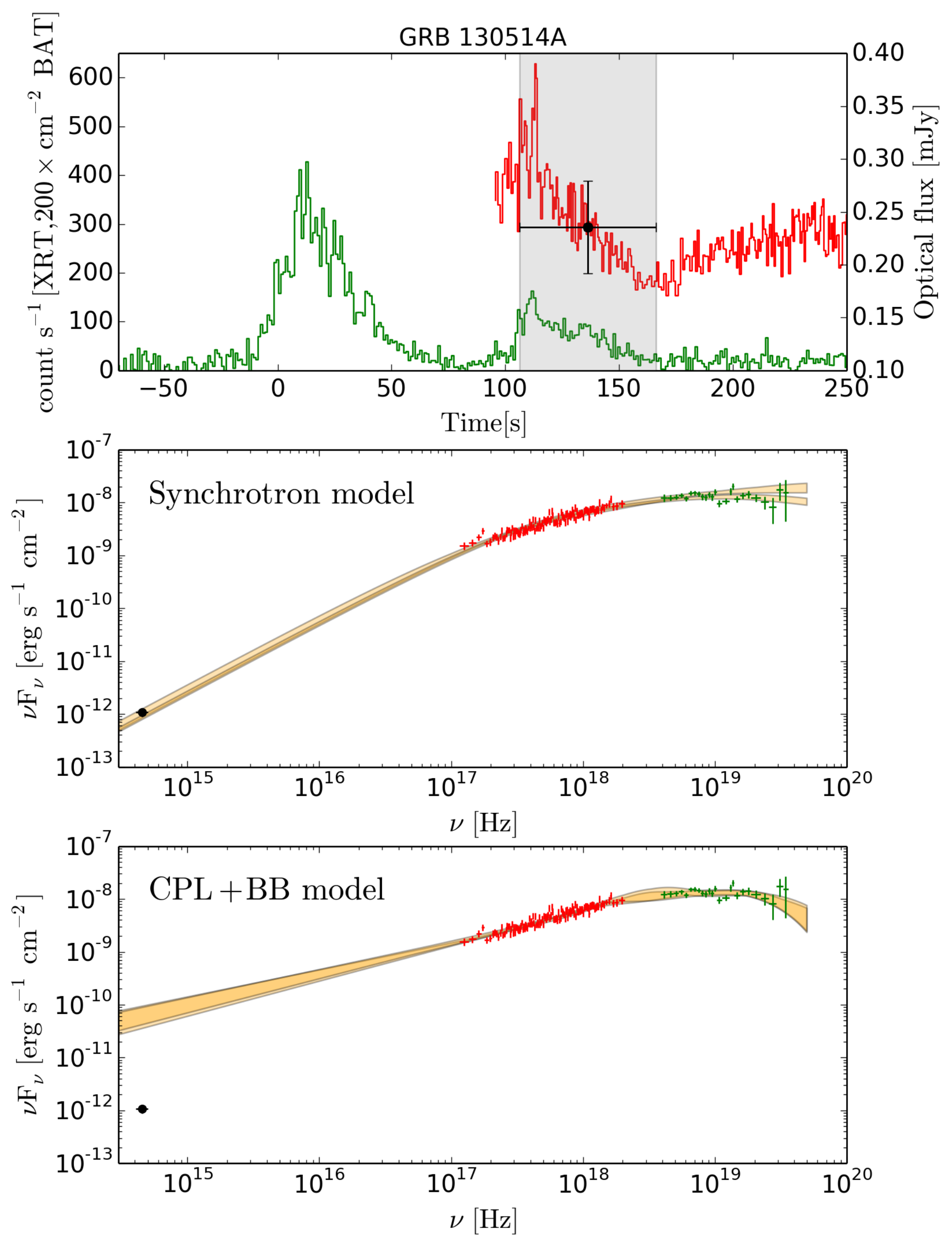}
   }
   \label{fig:singleexp2}
\end{center}
\end{figure*}
\begin{figure*}[ht!]
\begin{center}
   {
    \includegraphics[width=0.48\textwidth]{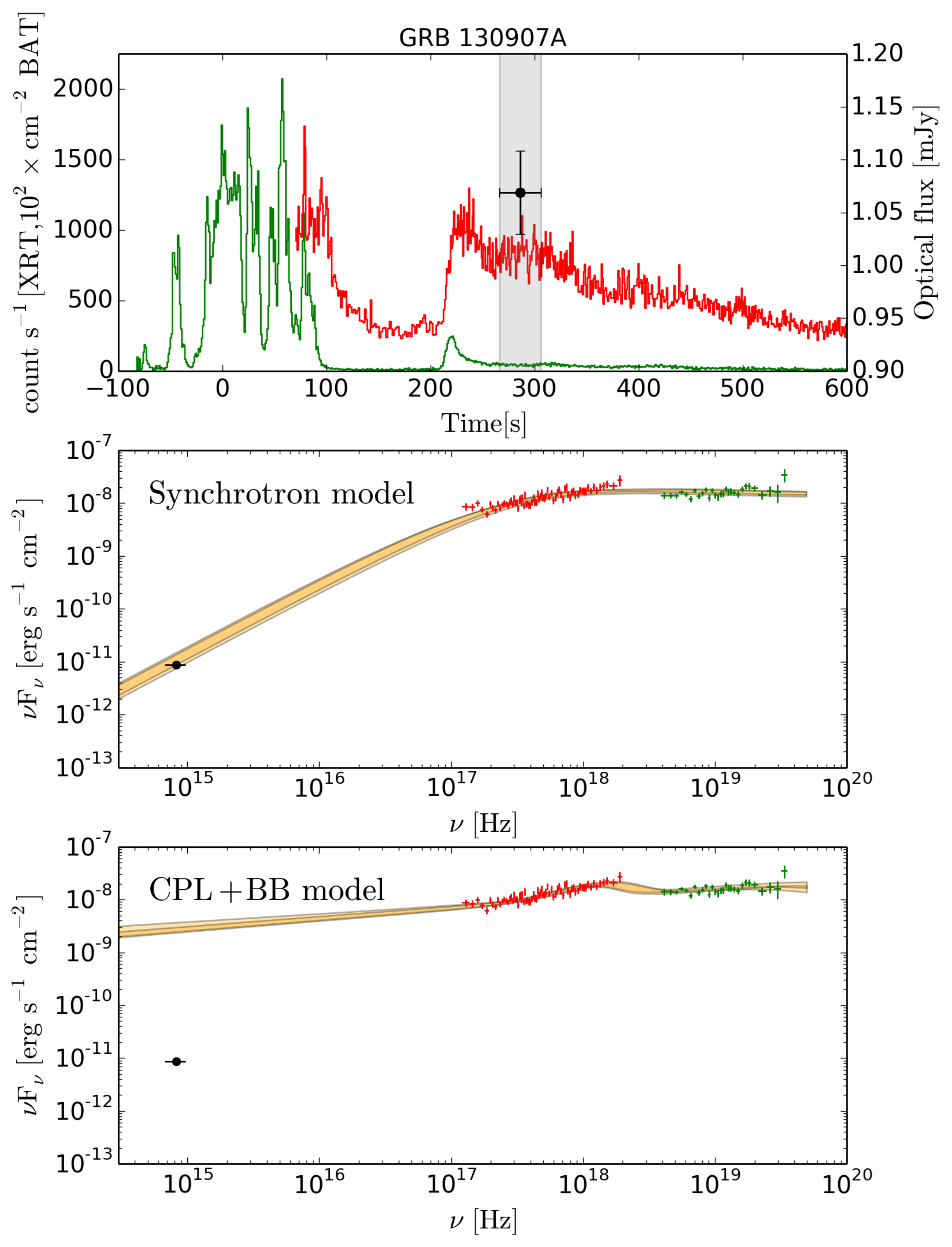}
   \hskip 0.5truecm
    \includegraphics[width=0.48\textwidth]{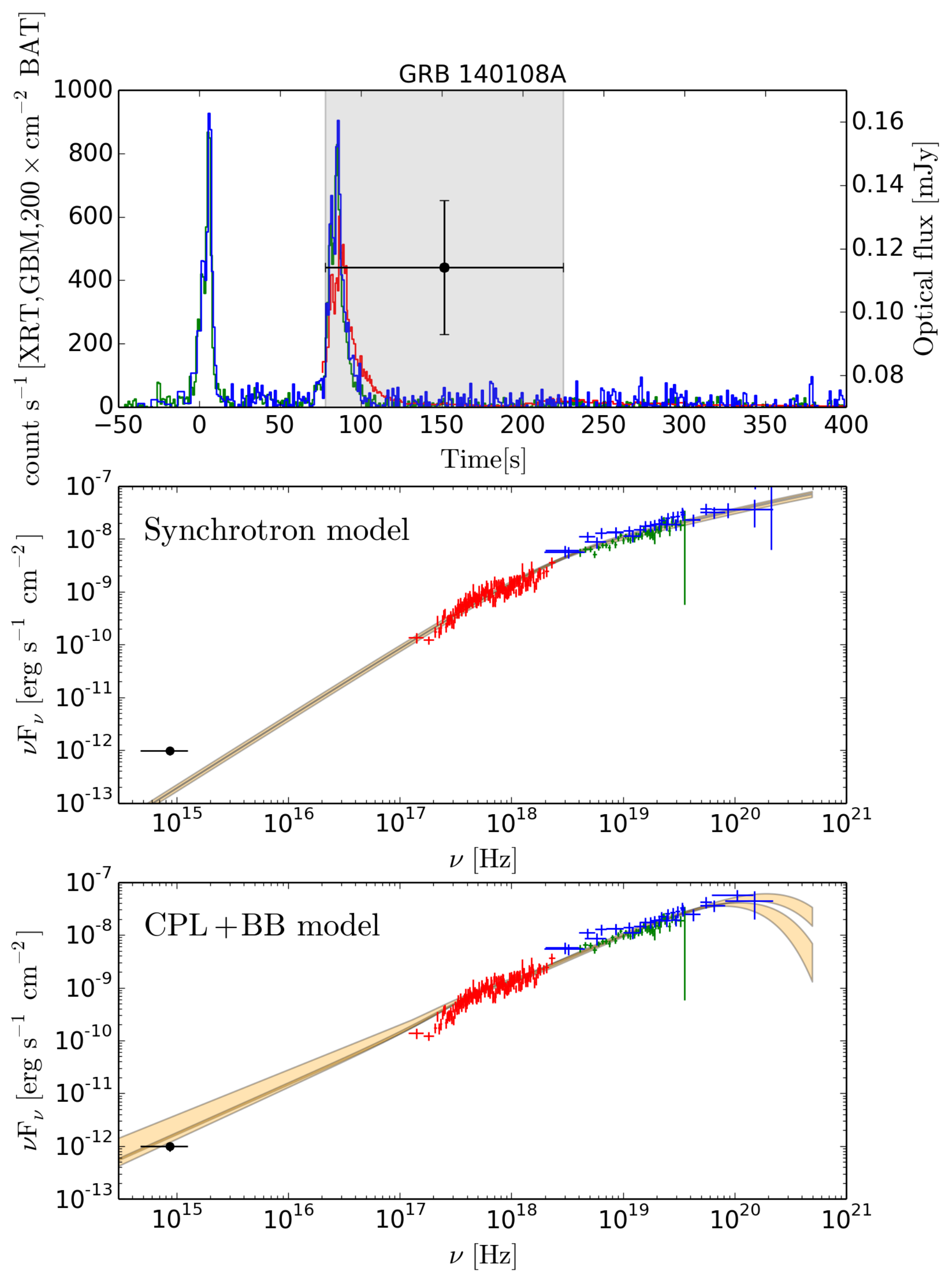}
    \vskip 0.8truecm
   \includegraphics[width=0.48\textwidth]{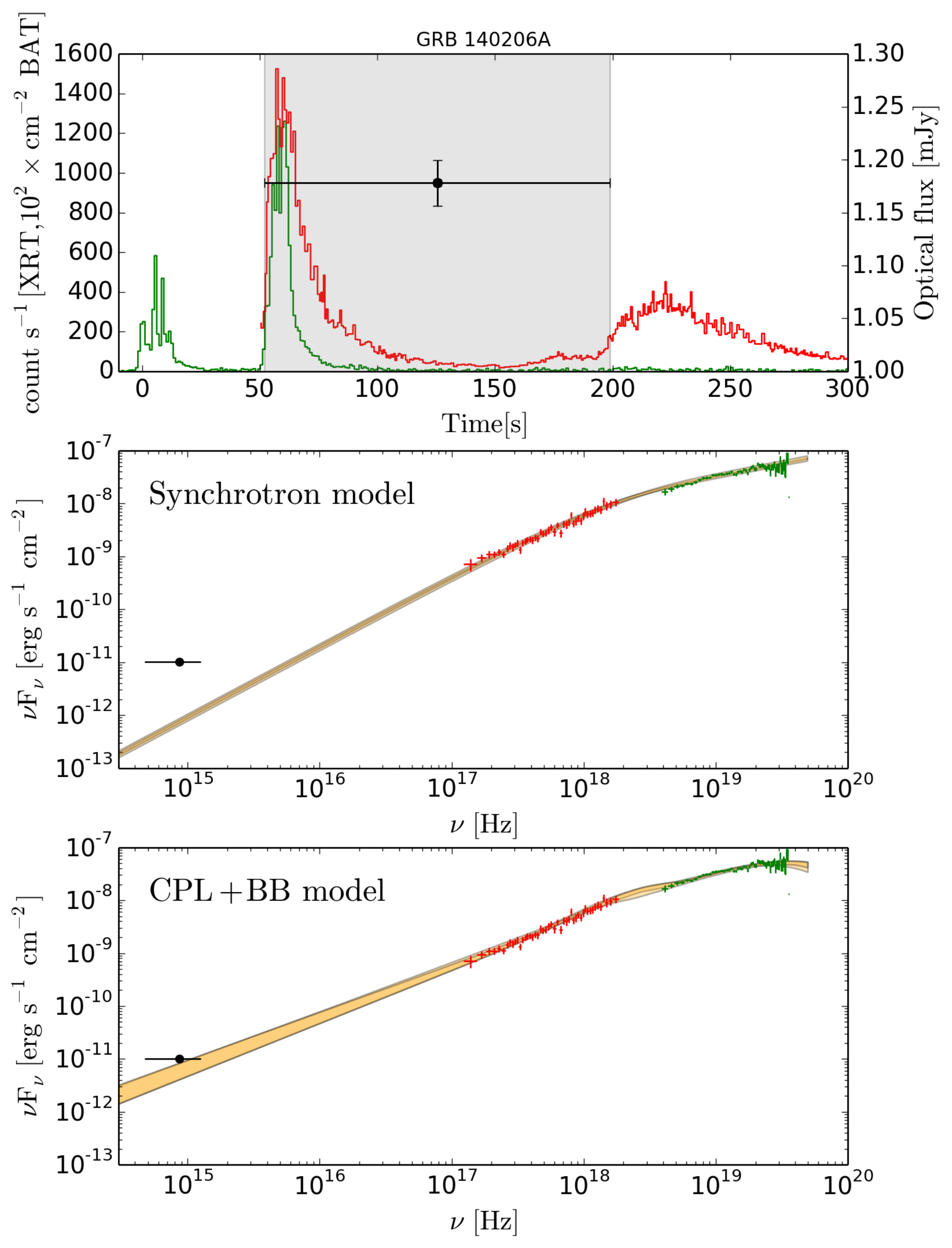}
   \hskip 0.5truecm
   \includegraphics[width=0.48\textwidth]{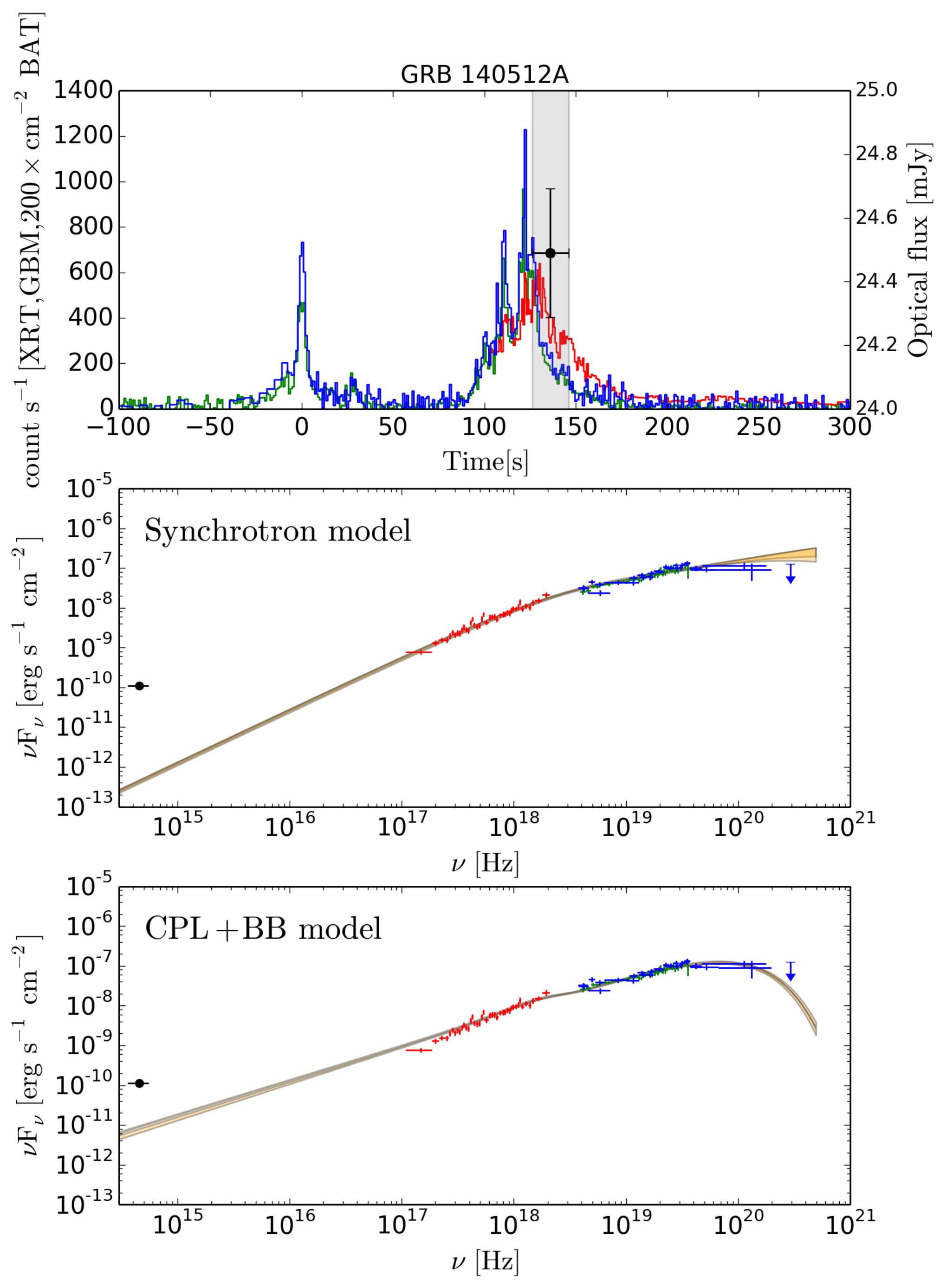}
   }
   \label{fig:singleexp3}
\end{center}
\end{figure*}

\begin{figure*}[ht!]
\begin{center}
   {
   \includegraphics[width=0.48\textwidth]{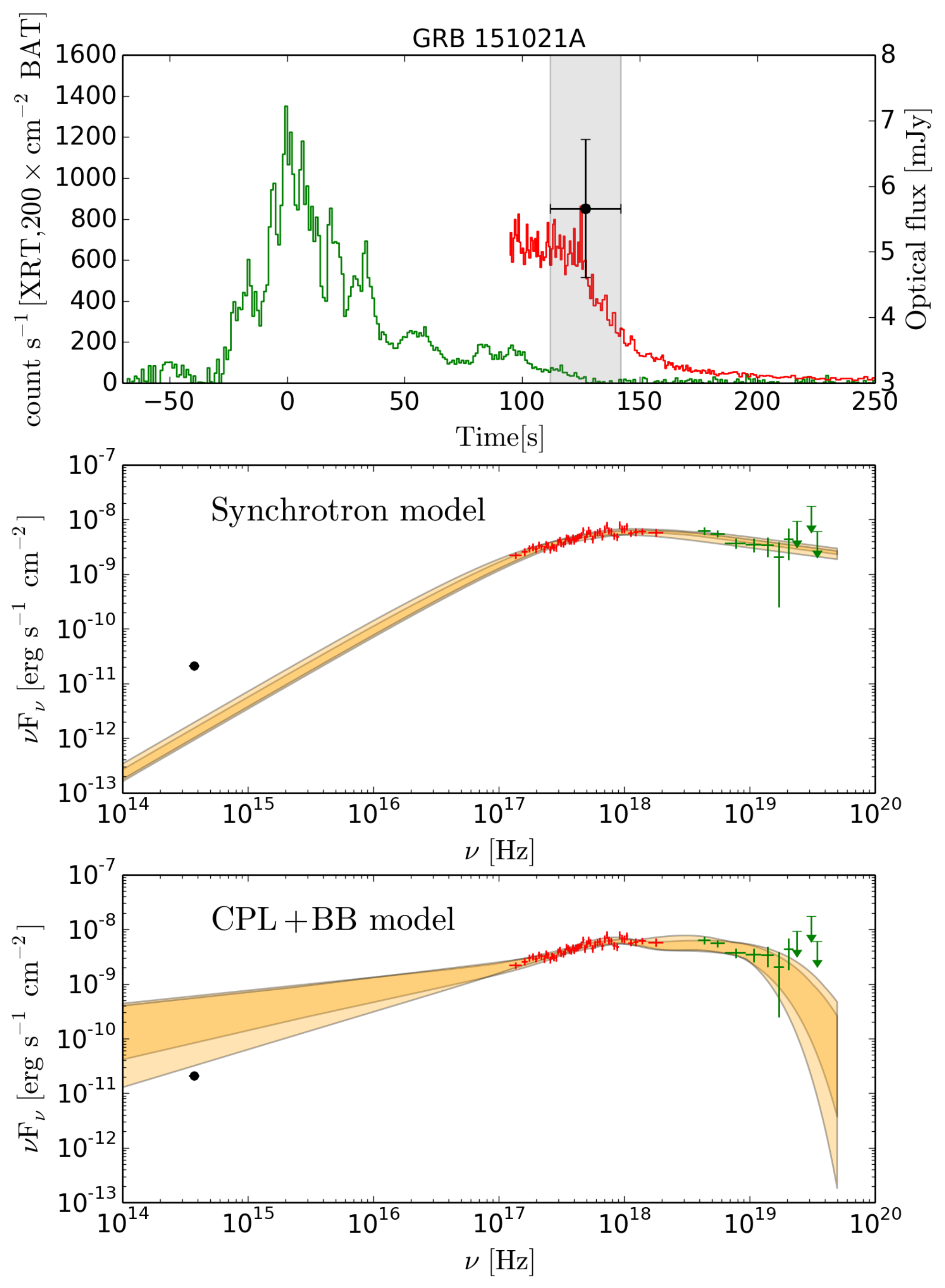}
   }
   \label{fig:singleexp4}
\end{center}
\end{figure*}

\end{appendix}
\end{document}